\documentclass[%
 reprint,prl, 
superscriptaddress,
nobibnotes,
 amsmath,amssymb,
 aps,
]{revtex4-2}
\usepackage[utf8]{inputenc}
\usepackage{graphicx}
\usepackage{dcolumn}
\usepackage{bm}
\usepackage{hyperref}
\usepackage[mathlines]{lineno}
\usepackage{amsmath} 

\relax

\begin{document}

\preprint{APS/123-QED}

\title{First Results from the Taiwan Axion Search Experiment with Haloscope at
 19.6$\,\mu\text{e\hspace{-.08em}V}$}

\author{Hsin~Chang}\affiliation{Department of Physics, National Central University, Taoyuan City 320317, Taiwan}
\author{Jing-Yang~Chang}\affiliation{Department of Physics, National Central University, Taoyuan City 320317, Taiwan} 
\author{Yi-Chieh~Chang}\affiliation{National Synchrotron Radiation Research Center, Hsinchu 300092, Taiwan} 
\author{Yu-Han~Chang}\affiliation{Department of Physics, National Chung Hsing University, Taichung City 402202, Taiwan}
\author{Yuan-Hann~Chang}\affiliation{Institute of Physics, Academia Sinica, Taipei City 115201, Taiwan}
\affiliation{Center for High Energy and High Field Physics, National Central University, Taoyuan City 320317, Taiwan}
\author{Chien-Han~Chen}\affiliation{Institute of Physics, Academia Sinica, Taipei City 115201, Taiwan} 
\author{Ching-Fang~Chen}\affiliation{Department of Physics, National Central University, Taoyuan City 320317, Taiwan}
\author{Kuan-Yu~Chen}\affiliation{Department of Physics, National Central University, Taoyuan City 320317, Taiwan} 
\author{Yung-Fu~Chen}\email[Correspondence to: ]{yfuchen@ncu.edu.tw}\affiliation{Department of Physics, National Central University, Taoyuan City 320317, Taiwan} 
\author{Wei-Yuan~Chiang}\affiliation{National Synchrotron Radiation Research Center, Hsinchu 300092, Taiwan}
\author{Wei-Chen~Chien}\affiliation{Department of Physics, National Chung Hsing University, Taichung City 402202, Taiwan}
\author{Hien~Thi~Doan}\affiliation{Institute of Physics, Academia Sinica, Taipei City 115201, Taiwan} 
\author{Wei-Cheng~Hung}\affiliation{Department of Physics, National Central University, Taoyuan City 320317, Taiwan}\affiliation{Institute of Physics, Academia Sinica, Taipei City 115201, Taiwan} 
\author{Watson~Kuo}\affiliation{Department of Physics, National Chung Hsing University, Taichung City 402202, Taiwan} 
\author{Shou-Bai~Lai}\affiliation{Department of Physics, National Central University, Taoyuan City 320317, Taiwan} 
\author{Han-Wen~Liu}\affiliation{Department of Physics, National Central University, Taoyuan City 320317, Taiwan} 
\author{Min-Wei~OuYang}\affiliation{Department of Physics, National Central University, Taoyuan City 320317, Taiwan}
\author{Ping-I~Wu}\affiliation{Department of Physics, National Central University, Taoyuan City 320317, Taiwan} 
\author{Shin-Shan~Yu}\email[Correspondence to: ]{syu@phy.ncu.edu.tw}\affiliation{Department of Physics, National Central University, Taoyuan City 320317, Taiwan}
\affiliation{Center for High Energy and High Field Physics, National Central University, Taoyuan City 320317, Taiwan}

\collaboration{TASEH Collaboration}

\begin{abstract}

 This Letter reports on the first results from the Taiwan Axion Search 
Experiment with Haloscope, a search for axions using a microwave cavity at 
frequencies between 4.70750 and 4.79815~GHz. Apart from the non-axion signals,
 no candidates with a significance more than 3.355 were found. The experiment 
excludes models with the axion-two-photon coupling 
$\left|g_{a\gamma\gamma}\right|\gtrsim 8.2\times 10^{-14}\,\text{Ge\hspace{-.08em}V}^{-1}$, a factor of eleven above the benchmark KSVZ model, reaching 
a sensitivity three orders of magnitude better than any existing limits 
 in the mass range $19.4687 < m_a < 19.8436 \,\mu\text{e\hspace{-.08em}V}$.  
 It is also the first time that a haloscope-type experiment places 
constraints on $g_{a\gamma\gamma}$ in this mass region.

\end{abstract}

\maketitle

Various astrophysical and cosmological observations indicate that dark matter 
(DM) exists and makes up 26.4\% of the total energy density of the 
universe~\cite{DMI,DMII,DMIII,DMIV,PDG}. One of the viable dark matter 
candidates is the axion, which arises from the spontaneous breaking of a new 
global U(1)$_\mathrm{PQ}$ symmetry~\cite{strongCPI} introduced by Peccei and 
Quinn to solve the strong CP problem~\cite{strongCPI,strongCPII,strongCPIII}. 
Axions are abundantly produced during the QCD phase transition in the early 
universe and may constitute the DM~\cite{ADDONI,ADDONII,ADDONIII,ADDONIV}. 
In the post-inflationary PQ symmetry breaking scenario, current calculations 
suggest a mass range of ${\cal O}(1–100)\,\mu\text{e\hspace{-.08em}V}$ for 
axions so that the cosmic axion density does not exceed the 
observed cold DM density~\cite{QCDCalI,QCDCalII,QCDCalIII,QCDCalIV,QCDCalV,QCDCalVI,QCDCalVII,QCDCalVIII,QCDCalIX,QCDCalX,QCDCalXI,QCDCalXII,QCDCalXIII}.

Axions could be detected and studied via their two-photon interaction, of 
which the strength is described by the coupling constant $g_{a\gamma\gamma}$. 
The detectors with the best sensitivities to axion DMs with a mass of 
$m_a\approx \,\mu\text{e\hspace{-.08em}V}$, as first proposed by 
Sikivie~\cite{SikivieI,SikivieII}, are haloscopes consisting of a microwave 
(MW) cavity immersed in a strong static magnetic field and operated at a 
cryogenic temperature. In the presence of an external magnetic field, the 
ambient oscillating axion field drives the cavity and they resonate when the 
frequencies of the electromagnetic modes in the cavity match the MW frequency 
$f$, where $f$ is set by the total energy of the axion: 
$hf=E_a=m_a c^2 + \frac{1}{2}m_a v^2$. The axion signal power is further 
delivered to the readout probe followed by a low-noise linear amplifier.

Several haloscope experiments have actively carried out axion searches. 
The most significant efforts are from the Axion Dark Matter eXperiment (ADMX),
 placing tight constraints on $g_{a\gamma\gamma}$ 
within the mass range of 
1.9--4.2$\,\mu\text{e\hspace{-.08em}V}$~\cite{ADMXI,ADMXII,ADMXIII,ADMXIV,ADMXV,ADMXVI,ADMXVII}. 
Others include the Haloscope at Yale Sensitive to Axion Cold dark matter 
(HAYSTAC)~\cite{HAYSTACIII,HAYSTACIV,HAYSTACI}, the Center 
for Axion and Precision Physics Research (CAPP)~\cite{CAPPII,CAPPIII,CAPPI}, 
and the QUest for AXions-$a\gamma$ (QUAX-$a\gamma$)~\cite{QUAX}. 
This Letter presents the first results of a search for axions in the mass 
range of 19.4687--19.8436$\,\mu\text{e\hspace{-.08em}V}$, from the Taiwan 
Axion Search Experiment with Haloscope (TASEH).

\begin{figure} 
  \centering
  \includegraphics[width=8.6cm]{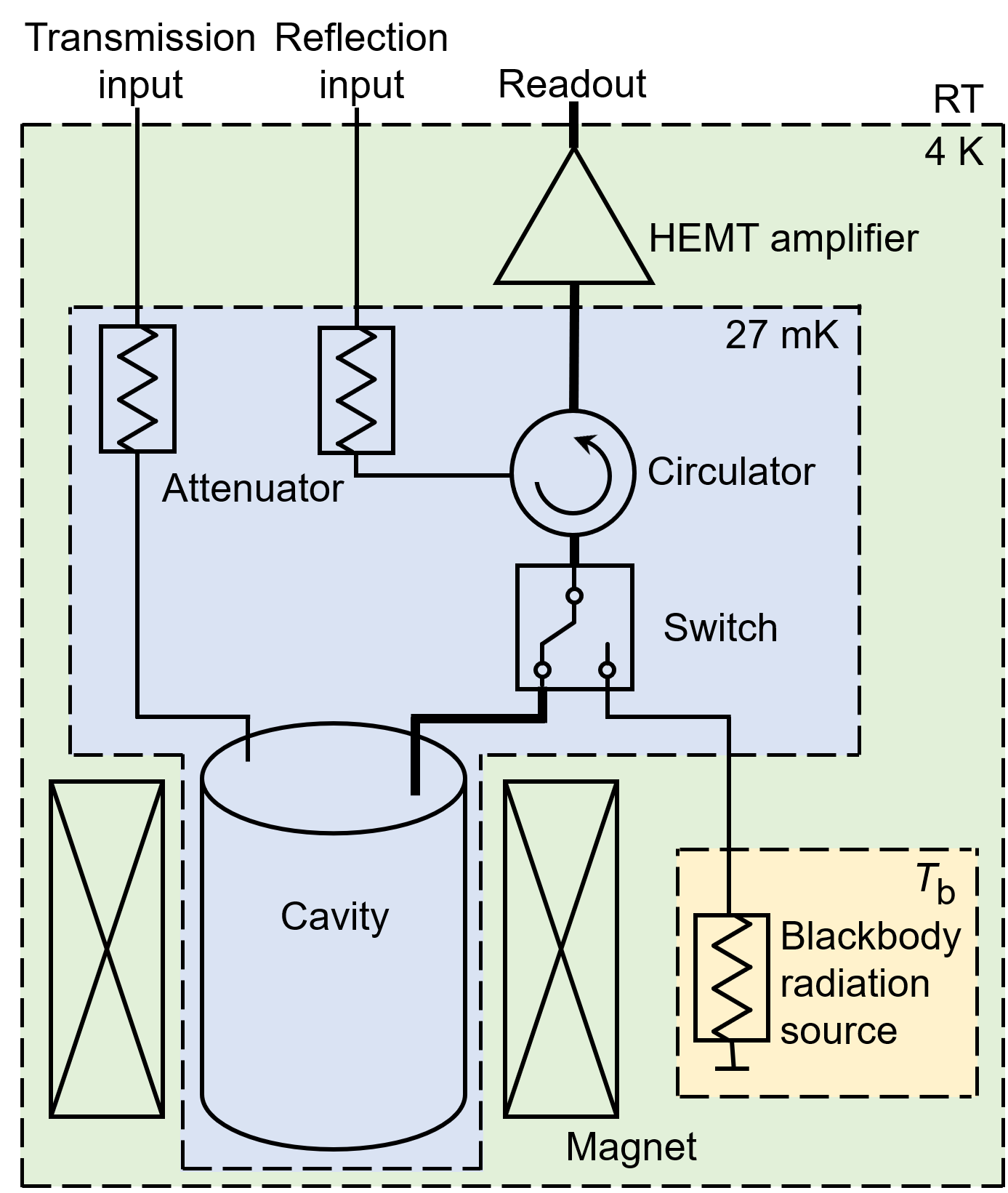}
  \caption{The simplified diagram of the TASEH apparatus.}
  \label{fig:TASEH}
\end{figure}

The detector of TASEH is located at the Department of Physics, National 
Central University, Taiwan and housed within a cryogen-free dilution 
refrigerator (DR) from BlueFors. Figure~\ref{fig:TASEH} shows a simplified 
diagram of the TASEH apparatus. An 8-Tesla superconducting solenoid with a 
bore diameter of 76~mm and a length of 240 mm is integrated with the DR.  
The DR has multiple flanges at different temperatures for the required 
cooling: 50~K, 4~K, still, and mixing flanges, among which the mixing flange 
could reach the lowest temperature at $\simeq20$~mK. During the data taking, 
the MW cavity with two coupling probes sits in the center of the magnet bore 
and is connected via holders to the mixing flange. The $0.234$-L cylindrical 
cavity, made of oxygen-free high-conductivity (OFHC) copper, has an inner 
radius of 2.5~cm and a height of 12~cm and is split into two halves along the 
axial direction to reduce the loss from the seam~\cite{CAPPCavity}. 
The resonant frequency can be tuned via the rotation of an off-axis OFHC 
copper tuning rod. The axion-photon conversion signal from the readout probe 
is directed to an impedance-matched amplification chain (thick lines in 
Fig.~\ref{fig:TASEH}). The first-stage amplifier is a low noise 
high-electron-mobility-transistor (HEMT) amplifier mounted on the 4~K flange. 
A circulator, anchored at the mixing flange, prevents thermal radiation from 
the HEMT amplifier back streaming to the cold cavity and then being reflected. 
 The signal is further amplified at room temperature via a three-stage 
post-amplifier, and down-converted to in-phase (I) and quadrature (Q) 
components and digitized by an analog-to-digital converter with a sampling 
rate of 2~MHz. Two heavily attenuated input lines support scattering parameter 
measurements of the cavity and transmit test signals. More details of the 
TASEH detector can be found in Ref.~\cite{TASEHInstrumentation}. 

The signal power extracted from a MW cavity on resonance is given 
by~\cite{AxionFormula,HAYSTACIII}:
\begin{equation}
P_s = \left(g_{a\gamma\gamma}^2\frac{\hbar^3c^3\rho_a}{m_a^2}\right)\times
\left(\omega_c\frac{1}{\mu_0}B_0^2VCQ_L\frac{\beta}{1+\beta}\right).
\label{eq:ps}
\end{equation}
The first set of parentheses contains $g_{a\gamma\gamma}$, $m_a$, physical 
constants, and the local dark-matter density 
$\rho_a=0.45\,\text{Ge\hspace{-.08em}V}/\mathrm{cm}^3$~\cite{[{}][{. 
Both $0.45\,\text{Ge\hspace{-.08em}V}/\mathrm{cm}^3$ 
(used by ADMX, HAYSTAC, CAPP, and QUAX-$a\gamma$) and 
$0.3\,\text{Ge\hspace{-.08em}V}/\mathrm{cm}^3$ (more commonly cited by the 
other direct DM search experiments) are consistent 
with the recent measurements.}]Read:2014qva,PDG}. 
For the QCD axions, $g_{a\gamma\gamma}$ is related to the axion mass $m_a$: 
\begin{equation}
 g_{a\gamma\gamma} = \left(\frac{g_{\gamma}\alpha}{\pi \Lambda^2}\right)m_a, 
\label{eq:grelation}
\end{equation}
where $g_{\gamma}$ is a dimensionless model-dependent parameter,    
with numerical values -0.97 and 0.36 
in the Kim-Shifman-Vainshtein-Zakharov (KSVZ)~\cite{KSVZI,KSVZII} and 
the Dine-Fischler-Srednicki-Zhitnitsky (DFSZ)~\cite{DFSZI,DFSZII} benchmark 
models, respectively. The symbol $\alpha$ is the fine-structure constant and  
$\Lambda=78~\,\text{Me\hspace{-.08em}V}$ is a scale parameter that can 
be derived from the mass and the decay constant of the pion and the ratio of 
the up to down quark masses. 
The second set of parentheses contains parameters related to the experimental 
setup: the angular resonant frequency of the cavity $\omega_c$, the vacuum 
permeability $\mu_0$, the nominal strength of the external magnetic field 
$B_0$, the effective volume of the cavity $V$, and the loaded quality factor 
of the cavity \(Q_L=Q_0/(1+\beta)\), where $Q_0$ is the unloaded, intrinsic 
quality factor and $\beta$ is the coupling coefficient which 
determines the amount of coupling of the signal to the receiver. The form 
factor $C$ is the normalized overlap of the electric field $\vec{\bm{E}}$, 
for a particular cavity resonant mode, with the external magnetic field 
$\vec{\bm{B}}$:
\begin{equation}
  C = \frac{\left[\int\left( \vec{\bm{B}}\cdot\vec{\bm{E}}\right) d^3\bm{x}\right]^2}{B_0^2V\int E^2 d^3\bm{x}}.
\label{eq:formfactor} 
\end{equation} 
The magnetic field $\vec{\bm{B}}$ in TASEH points mostly along the axial
direction of the cavity. For cylindrical cavities, the largest form factor is 
from the TM$_{010}$ mode. The expected signal power derived from the 
experimental parameters of TASEH is $P_s\simeq 1.4\times10^{-24}$~W for a 
KSVZ axion with a mass of 19.6$\,\mu\text{e\hspace{-.08em}V}$. 

In the haloscope experiments, the figure of merit that determines the design 
of the experimental setup is the signal-to-noise ratio
(SNR), i.e. the ratio of the signal power $P_s$ to the fluctuation in 
the averaged noise power spectrum $\sigma_n$~\cite{Dicke}, given by:  
\begin{equation}
   \text{SNR}  =  \frac{P_s}{\sigma_n}=  \frac{P_s}{k_B T_\text{sys}}\sqrt{\frac{t}{\Delta f}},
 \label{eq:SNR}
\end{equation}  
where $T_\text{sys}$ is the system noise temperature, an effective temperature
 associated with the total noise of the system, $t$ is the data integration 
time, and $\Delta f$ is the resolution bandwidth. Here, one assumes that all 
the axion signal power falls within $\Delta f$.

The system noise temperature $T_\text{sys}$ has three major components:
\begin{equation}
 T_\text{sys} = \Tilde{T}_\text{mx} + \left(\Tilde{T}_\text{c}-\Tilde{T}_\text{mx}\right)L(\omega) + T_\text{a},
\label{eq:pn}
\end{equation}
where $\omega$ is the angular frequency.
The last term $T_\text{a}$ is the effective temperature of the
noise added by the receiver (mainly from the first-stage amplifier).
The sum of the first two terms is equivalent to the sum of the reflection of 
the incoming noise from the attenuator anchored to the mixing flange 
(Fig.~\ref{fig:TASEH}) and the transmission of the noise from the cavity 
body itself. The symbol 
$\Tilde{T}_i=\left(\frac{1}{e^{\left.\hbar\omega\middle/k_BT_i\right.}-1} + \frac{1}{2}\right)\left. \hbar\omega\middle/k_B\right.$ refers to
the effective temperature due to the blackbody radiation at a physical 
temperature $T_i$ and the vacuum fluctuation. The difference of the effective 
temperatures $\Tilde{T}_\text{c}-\Tilde{T}_\text{mx}$ is modulated by a 
Lorentzian function $L(\omega)$. If the physical temperatures of the cavity 
$T_\text{c}$ and of the mixing flange $T_\text{mx}$ are identical, the thermal
 noise spectrum from the cavity is flat. The derivation of the first two terms
 in Eq.~\eqref{eq:pn} can be found in Ref.~\cite{TASEHAnalysis}.

The calibration for the amplification chain is performed by connecting the HEMT
 to a blackbody radiation source (Fig.~\ref{fig:TASEH}) instead of the cavity 
via a cryogenic switch. Various values of input currents are sent to a nearby 
resistor heater to change its temperature $T_{\text b}$ monitored by a 
thermometer. The output power is fitted to a first-order polynomial, as a 
function of the source temperature, to extract the overall gain and added 
noise $T_\text{a}$.

The data for the analysis presented here were collected by TASEH from October 
13, 2021 to November 15, 2021, and are termed as the CD102 data, where CD 
stands for ``cool down''. The CD102 data cover the frequency range of 
4.70750--4.79815~GHz. In this Letter, most of the frequencies in unit of GHz 
are quoted with five decimal places as the absolute accuracy of frequency is 
$\approx 10$~kHz. It shall be noted that the frequency resolution is 1~kHz.  
During the CD102 data run, the temperature of the cavity stayed at 
$T_\text{c}\simeq155$~mK, higher with respect to the mixing flange 
$T_\text{mx}\simeq27$~mK; it is believed that the cavity had an unexpected 
thermal contact with the radiation shield in the DR. As a result, 
$\Tilde{T}_\text{c}$ and $\Tilde{T}_\text{mx}$ are 0.18~K and 0.11~K, 
respectively. The form factor $C$ for the TM$_{010}$ mode varies from 0.60 to 
0.61 over the operational frequency range. The intrinsic quality factor $Q_0$ 
at the cryogenic temperature is $\simeq 60700$. The insertion depth of the 
readout probe is set for $\beta\simeq2$ since this value, for a given amount 
of time and a fixed value of SNR, maximizes the frequency coverage. 
In this case, the cavity line width, $\left.\omega_c\middle/2\pi Q_L\right.$, 
is about 240~kHz. 

The calibration was carried out before, during, and after the data taking,
which showed that the performance of the system was stable over time. 
The added noise $T_\text{a}$ obtained from the calibration is 
about $1.9-2.2$~K, with a frequency dependence. In CD102, 
there were 837 resonant-frequency steps in total, with a frequency difference 
of $\Delta f_\text{s}=95-115$~kHz between the steps. The value of 
$\Delta f_\text{s}$ was kept within 10\% of 105~kHz ($\lesssim$ half of the 
cavity line width) rather than a fixed value, such that the rotation angle of 
the tuning rod did not need to be fine-tuned and the operation time could be 
minimized. A 10\% variation of $\Delta f_\text{s}$ is found to have no impact 
on the $g_{a\gamma\gamma}$ limits. Each resonant-frequency step is denoted as 
a ``scan'' and the data integration time was about 32-42 minutes. 
The variation of the integration time was introduced to compensate 
the frequency dependence of the added noise. 

The analysis of the CD102 data follows the procedure similar to that 
developed by the HAYSTAC experiment~\cite{HAYSTACII} and the details are 
described in Ref.~\cite{TASEHAnalysis}. The fast Fourier transform 
algorithm is performed on the IQ time series data to obtain the 
frequency-domain power spectrum. The Savitzky-Golay (SG) 
filter~\cite{SGFilter} is applied to model the Lorentzian structure of 
the background caused by the temperature difference between the cavity and 
the mixing flange [Eq.~\eqref{eq:pn}] and to obtain the 
average noise power. 
Deviations from the average noise power are compared with the 
uncertainty on the averaged power spectrum, which defines the 
observed SNR. All the spectra from different 
frequency scans, particularly for the frequency bins that appear in 
multiple spectra, are combined with a weighting algorithm. 
In order to maximize the SNR, a running window of 
five consecutive bins in the combined spectrum is applied and the five bins 
within each window are merged to construct a final spectrum. 
The five frequency bins correspond to the 5-kHz axion signal line width,  
assuming a standard Maxwellian axion line shape with a velocity 
variance $\left<v^2\right>=$(270~km/s)$^2$~\cite{HAYSTACII}. 
This line shape is also used when defining the maximum likelihood 
weights for merging.

After the merging, 22 candidates with an SNR greater than 3.355 were found and 
a rescan was performed to check if they were real signals  
or statistical fluctuations.        
Among them, 20 candidates were from the fluctuations because they were gone 
after a few rescans. 
The remaining two candidates, in the frequency ranges of 
4.71017 -- 4.71019~GHz and 4.74730 -- 4.74738~GHz, are not 
considered axion signal candidates for the following reasons. 
The signal in the second frequency range was detected via a portable antenna 
outside the DR and found to come from the instrument control computer in the 
laboratory, while the signal in the first frequency range was not detected 
outside the DR but still present after turning off the external magnetic 
field. No limits are placed for the above two frequency ranges.

\begin{figure*} 
  \centering
  \includegraphics[width=12.9cm]{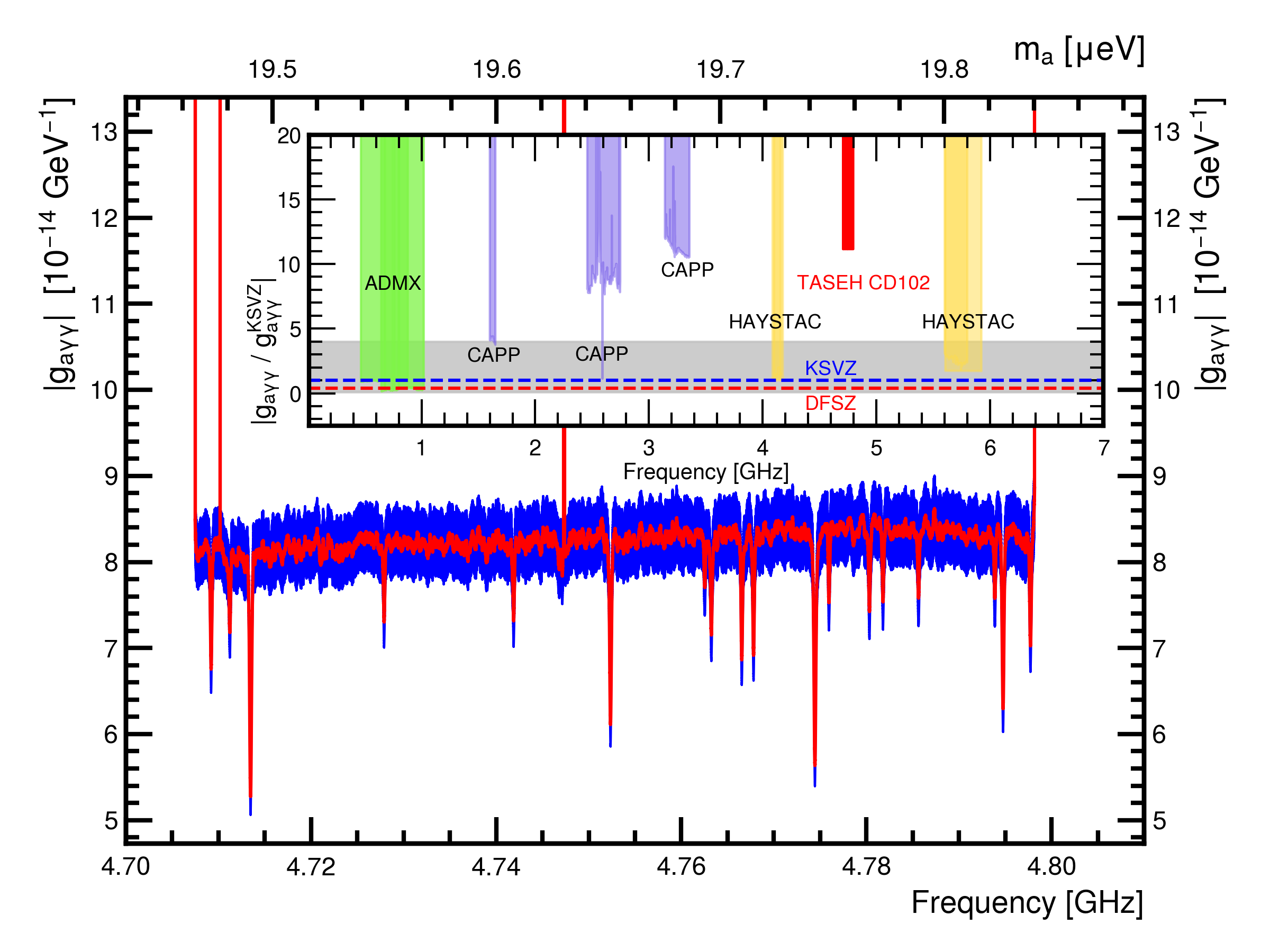}
  \caption{The 95\% C.L. limits on $\left|g_{a\gamma\gamma}\right|$ and the 
 ratio of the average limit with respect to the KSVZ benchmark value from the 
 CD102 data (red band in the inset). The results from the previous searches 
 performed by the ADMX, CAPP, and HAYSTAC Collaborations are also shown 
 (inset). The blue error band indicates the systematic uncertainties. The gray
 band in the inset shows the allowed region of 
 $\left|g_{a\gamma\gamma}\right|$ vs. $m_a$ from various QCD axion models, 
 while the blue and red dashed lines are the values predicted by the KSVZ and 
 DFSZ benchmark models, respectively.
 }

  \label{fig:gaggall}
\end{figure*}

Since no candidates were found after the rescan, the upper limits at 95\% 
confidence level (C.L.) on $\left|g_{a\gamma\gamma}\right|$ are derived by 
setting the maximum SNR equal to five, with the assumption that axions make up 
100\% of the local dark matter density.  
Figure~\ref{fig:gaggall} shows the $\left|g_{a\gamma\gamma}\right|$ limits of 
TASEH and the ratios relative to the KSVZ benchmark value, together with those 
from the previous searches. The limits on $\left|g_{a\gamma\gamma}\right|$ 
range from $5.3\times 10^{-14}\,\text{Ge\hspace{-.08em}V}^{-1}$ to 
$8.9\times 10^{-14}\,\text{Ge\hspace{-.08em}V}^{-1}$, with an average 
value of $8.2\times 10^{-14}\,\text{Ge\hspace{-.08em}V}^{-1}$; the lowest 
value comes from the frequency bins with additional eight times more data from 
the rescans, while the highest value comes from the frequency bins near the 
boundaries of the spectrum. Overall the total relative systematic uncertainty 
is $\approx 4.6\%$, coming from the uncertainties on the loaded quality factor 
$Q_L$, the coupling coefficient $\beta$, the added noise temperature 
$T_\text{a}$, the effect of the misalignment between the true axion frequency 
and the lower boundaries of the frequency bins, and the variation of the 
SG-filter parameters.

The analysis that merges bins without assuming a signal line shape results in 
$\approx5.5$\% larger values on the $\left|g_{a\gamma\gamma}\right|$ limits. 
If a Gaussian signal line shape with an FWHM of 2.5~kHz is assumed instead, 
the limits will be $\approx3.8$\% smaller than the central results. If the 
$\left|g_{a\gamma\gamma}\right|$ limits are derived from the observed SNR as 
described in the ADMX paper~\cite{ADMXVIII}, rather than using the 5$\sigma$ 
target SNR, the average limit on $\left|g_{a\gamma\gamma}\right|$ will 
be $\approx 4.9\times 10^{-14}\,\text{Ge\hspace{-.08em}V}^{-1}$.

After the collection of the CD102 data, synthetic axion signals were injected 
into the cavity via the transmission input line (Fig.~\ref{fig:TASEH}) and 
read out via the same amplification chain. The procedure to generate axion-like
 signals is summarized in Ref.~\cite{TASEHInstrumentation} and the analysis of
 the synthetic axion data is described in Ref.~\cite{TASEHAnalysis}. 
The analysis results demonstrate the capability of the experimental setup and 
the analysis strategy to discover an axion signal with 
$\left|g_{a\gamma\gamma}\right|\approx {\cal O}\left(10\left|g_{a\gamma\gamma}^\text{KSVZ}\right|\right)$.

In summary, a search for axions in the mass range 
$19.4687 < m_a < 19.8436 \,\mu\text{e\hspace{-.08em}V}$ was performed by the 
TASEH Collaboration. Apart from the non-axion signals, no candidates with a 
significance more than 3.355 were found. The experiment excludes models with 
the axion-two-photon coupling $\left|g_{a\gamma\gamma}\right|\gtrsim 8.2\times 10^{-14}\,\text{Ge\hspace{-.08em}V}^{-1}$ at 95\% C.L.,
 a factor of eleven above the benchmark KSVZ model. The sensitivity on 
$\left|g_{a\gamma\gamma}\right|$ reached by TASEH is three orders of magnitude 
better than the existing limits in the same mass range.  
It is also the first time that a haloscope-type experiment places 
constraints in this mass region. The target of TASEH is to search for axions 
in the mass range of 16.5--20.7$\,\mu\text{e\hspace{-.08em}V}$ 
corresponding to a frequency range of 4--5~GHz. 
With the upcoming upgrades of the experimental setup and several years of data 
taking, TASEH is expected to probe the QCD axion band in the target mass range.

\begin{acknowledgments}
We thank Chao-Lin Kuo for his help to initiate this project as well as
discussions on the microwave cavity design, Gray Rybka and Nicole Crisosto
for their introduction of the ADMX experimental
setup and analysis, Anson Hook for the discussions and the review of the
axion theory, and Jiunn-Wei Chen, Cheng-Wei Chiang, Cheng-Pang Liu, and
Asuka Ito for the discussions of future improvements in axion searches. 
We acknowledge support of microwave test and measurement equipment from the 
National Chung-Shan Institute of Science and Technology. 
  The work of the TASEH Collaboration was funded by
the Ministry of Science and Technology (MoST) of Taiwan with grant numbers
MoST-109-2123-M-001-002, MoST-110-2123-M-001-006, MoST-110-2112-M-213-018,
MoST-110-2628-M-008-003-MY3,
and MoST-109-2112-M-008-013-MY3, and by the Institute of Physics, Academia
Sinica.
\end{acknowledgments}

\providecommand{\noopsort}[1]{}\providecommand{\singleletter}[1]{#1}%


\begin{thebibliography}{54}%
\makeatletter
\providecommand \@ifxundefined [1]{%
 \@ifx{#1\undefined}
}%
\providecommand \@ifnum [1]{%
 \ifnum #1\expandafter \@firstoftwo
 \else \expandafter \@secondoftwo
 \fi
}%
\providecommand \@ifx [1]{%
 \ifx #1\expandafter \@firstoftwo
 \else \expandafter \@secondoftwo
 \fi
}%
\providecommand \natexlab [1]{#1}%
\providecommand \enquote  [1]{``#1''}%
\providecommand \bibnamefont  [1]{#1}%
\providecommand \bibfnamefont [1]{#1}%
\providecommand \citenamefont [1]{#1}%
\providecommand \href@noop [0]{\@secondoftwo}%
\providecommand \href [0]{\begingroup \@sanitize@url \@href}%
\providecommand \@href[1]{\@@startlink{#1}\@@href}%
\providecommand \@@href[1]{\endgroup#1\@@endlink}%
\providecommand \@sanitize@url [0]{\catcode `\\12\catcode `\$12\catcode
  `\&12\catcode `\#12\catcode `\^12\catcode `\_12\catcode `\%12\relax}%
\providecommand \@@startlink[1]{}%
\providecommand \@@endlink[0]{}%
\providecommand \url  [0]{\begingroup\@sanitize@url \@url }%
\providecommand \@url [1]{\endgroup\@href {#1}{\urlprefix }}%
\providecommand \urlprefix  [0]{URL }%
\providecommand \Eprint [0]{\href }%
\providecommand \doibase [0]{https://doi.org/}%
\providecommand \selectlanguage [0]{\@gobble}%
\providecommand \bibinfo  [0]{\@secondoftwo}%
\providecommand \bibfield  [0]{\@secondoftwo}%
\providecommand \translation [1]{[#1]}%
\providecommand \BibitemOpen [0]{}%
\providecommand \bibitemStop [0]{}%
\providecommand \bibitemNoStop [0]{.\EOS\space}%
\providecommand \EOS [0]{\spacefactor3000\relax}%
\providecommand \BibitemShut  [1]{\csname bibitem#1\endcsname}%
\let\auto@bib@innerbib\@empty
\bibitem [{\citenamefont {Gaitskell}(2004)}]{DMI}%
  \BibitemOpen
  \bibfield  {author} {\bibinfo {author} {\bibfnamefont {R.~J.}\ \bibnamefont
  {Gaitskell}},\ }\href {https://doi.org/10.1146/annurev.nucl.54.070103.181244}
  {\bibfield  {journal} {\bibinfo  {journal} {Ann. Rev. Nucl. Part. Sci.}\
  }\textbf {\bibinfo {volume} {54}},\ \bibinfo {pages} {315} (\bibinfo {year}
  {2004})}\BibitemShut {NoStop}%
\bibitem [{\citenamefont {Trimble}(1987)}]{DMII}%
  \BibitemOpen
  \bibfield  {author} {\bibinfo {author} {\bibfnamefont {V.}~\bibnamefont
  {Trimble}},\ }\href {https://doi.org/10.1146/annurev.aa.25.090187.002233}
  {\bibfield  {journal} {\bibinfo  {journal} {Ann. Rev. Astron. Astrophys.}\
  }\textbf {\bibinfo {volume} {25}},\ \bibinfo {pages} {425} (\bibinfo {year}
  {1987})}\BibitemShut {NoStop}%
\bibitem [{\citenamefont {Porter}\ \emph {et~al.}(2011)\citenamefont {Porter},
  \citenamefont {Johnson},\ and\ \citenamefont {Graham}}]{DMIII}%
  \BibitemOpen
  \bibfield  {author} {\bibinfo {author} {\bibfnamefont {T.~A.}\ \bibnamefont
  {Porter}}, \bibinfo {author} {\bibfnamefont {R.~P.}\ \bibnamefont
  {Johnson}},\ and\ \bibinfo {author} {\bibfnamefont {P.~W.}\ \bibnamefont
  {Graham}},\ }\href {https://doi.org/10.1146/annurev-astro-081710-102528}
  {\bibfield  {journal} {\bibinfo  {journal} {Ann. Rev. Astron. Astrophys.}\
  }\textbf {\bibinfo {volume} {49}},\ \bibinfo {pages} {155} (\bibinfo {year}
  {2011})}\BibitemShut {NoStop}%
\bibitem [{\citenamefont {Aghanim}\ \emph {et~al.}(2020)\citenamefont {Aghanim}
  \emph {et~al.}}]{DMIV}%
  \BibitemOpen
  \bibfield  {author} {\bibinfo {author} {\bibfnamefont {N.}~\bibnamefont
  {Aghanim}} \emph {et~al.} (\bibinfo {collaboration} {Planck}),\ }\href
  {https://doi.org/10.1051/0004-6361/201833910} {\bibfield  {journal} {\bibinfo
   {journal} {Astron. Astrophys.}\ }\textbf {\bibinfo {volume} {641}},\
  \bibinfo {pages} {A6} (\bibinfo {year} {2020})},\ \bibinfo {note} {[Erratum:
  Astron.Astrophys. 652, C4 (2021)]}\BibitemShut {NoStop}%
\bibitem [{\citenamefont {Zyla}\ \emph {et~al.}(2021)\citenamefont {Zyla} \emph
  {et~al.}}]{PDG}%
  \BibitemOpen
  \bibfield  {author} {\bibinfo {author} {\bibfnamefont {P.~A.}\ \bibnamefont
  {Zyla}} \emph {et~al.} (\bibinfo {collaboration} {Particle Data Group}),\
  }\href {https://doi.org/10.1093/ptep/ptaa104} {\bibfield  {journal} {\bibinfo
   {journal} {PTEP}\ }\textbf {\bibinfo {volume} {2020}},\ \bibinfo {pages}
  {083C01} (\bibinfo {year} {2021})}\BibitemShut {NoStop}%
\bibitem [{\citenamefont {Peccei}\ and\ \citenamefont
  {Quinn}(1977)}]{strongCPI}%
  \BibitemOpen
  \bibfield  {author} {\bibinfo {author} {\bibfnamefont {R.~D.}\ \bibnamefont
  {Peccei}}\ and\ \bibinfo {author} {\bibfnamefont {H.~R.}\ \bibnamefont
  {Quinn}},\ }\href {https://doi.org/10.1103/PhysRevLett.38.1440} {\bibfield
  {journal} {\bibinfo  {journal} {Phys. Rev. Lett.}\ }\textbf {\bibinfo
  {volume} {38}},\ \bibinfo {pages} {1440} (\bibinfo {year}
  {1977})}\BibitemShut {NoStop}%
\bibitem [{\citenamefont {Weinberg}(1978)}]{strongCPII}%
  \BibitemOpen
  \bibfield  {author} {\bibinfo {author} {\bibfnamefont {S.}~\bibnamefont
  {Weinberg}},\ }\href {https://doi.org/10.1103/PhysRevLett.40.223} {\bibfield
  {journal} {\bibinfo  {journal} {Phys. Rev. Lett.}\ }\textbf {\bibinfo
  {volume} {40}},\ \bibinfo {pages} {223} (\bibinfo {year} {1978})}\BibitemShut
  {NoStop}%
\bibitem [{\citenamefont {Wilczek}(1978)}]{strongCPIII}%
  \BibitemOpen
  \bibfield  {author} {\bibinfo {author} {\bibfnamefont {F.}~\bibnamefont
  {Wilczek}},\ }\href {https://doi.org/10.1103/PhysRevLett.40.279} {\bibfield
  {journal} {\bibinfo  {journal} {Phys. Rev. Lett.}\ }\textbf {\bibinfo
  {volume} {40}},\ \bibinfo {pages} {279} (\bibinfo {year} {1978})}\BibitemShut
  {NoStop}%
\bibitem [{\citenamefont {Preskill}\ \emph {et~al.}(1983)\citenamefont
  {Preskill}, \citenamefont {Wise},\ and\ \citenamefont {Wilczek}}]{ADDONI}%
  \BibitemOpen
  \bibfield  {author} {\bibinfo {author} {\bibfnamefont {J.}~\bibnamefont
  {Preskill}}, \bibinfo {author} {\bibfnamefont {M.~B.}\ \bibnamefont {Wise}},\
  and\ \bibinfo {author} {\bibfnamefont {F.}~\bibnamefont {Wilczek}},\ }\href
  {https://doi.org/https://doi.org/10.1016/0370-2693(83)90637-8} {\bibfield
  {journal} {\bibinfo  {journal} {Physics Letters B}\ }\textbf {\bibinfo
  {volume} {120}},\ \bibinfo {pages} {127} (\bibinfo {year}
  {1983})}\BibitemShut {NoStop}%
\bibitem [{\citenamefont {Abbott}\ and\ \citenamefont
  {Sikivie}(1983)}]{ADDONII}%
  \BibitemOpen
  \bibfield  {author} {\bibinfo {author} {\bibfnamefont {L.}~\bibnamefont
  {Abbott}}\ and\ \bibinfo {author} {\bibfnamefont {P.}~\bibnamefont
  {Sikivie}},\ }\href
  {https://doi.org/https://doi.org/10.1016/0370-2693(83)90638-X} {\bibfield
  {journal} {\bibinfo  {journal} {Physics Letters B}\ }\textbf {\bibinfo
  {volume} {120}},\ \bibinfo {pages} {133} (\bibinfo {year}
  {1983})}\BibitemShut {NoStop}%
\bibitem [{\citenamefont {Dine}\ and\ \citenamefont
  {Fischler}(1983)}]{ADDONIII}%
  \BibitemOpen
  \bibfield  {author} {\bibinfo {author} {\bibfnamefont {M.}~\bibnamefont
  {Dine}}\ and\ \bibinfo {author} {\bibfnamefont {W.}~\bibnamefont
  {Fischler}},\ }\href
  {https://doi.org/https://doi.org/10.1016/0370-2693(83)90639-1} {\bibfield
  {journal} {\bibinfo  {journal} {Physics Letters B}\ }\textbf {\bibinfo
  {volume} {120}},\ \bibinfo {pages} {137} (\bibinfo {year}
  {1983})}\BibitemShut {NoStop}%
\bibitem [{\citenamefont {Ipser}\ and\ \citenamefont
  {Sikivie}(1983)}]{ADDONIV}%
  \BibitemOpen
  \bibfield  {author} {\bibinfo {author} {\bibfnamefont {J.}~\bibnamefont
  {Ipser}}\ and\ \bibinfo {author} {\bibfnamefont {P.}~\bibnamefont
  {Sikivie}},\ }\href {https://doi.org/10.1103/PhysRevLett.50.925} {\bibfield
  {journal} {\bibinfo  {journal} {Phys. Rev. Lett.}\ }\textbf {\bibinfo
  {volume} {50}},\ \bibinfo {pages} {925} (\bibinfo {year} {1983})}\BibitemShut
  {NoStop}%
\bibitem [{\citenamefont {Borsanyi}\ \emph {et~al.}(2016)\citenamefont
  {Borsanyi}, \citenamefont {Fodor}, \citenamefont {Guenther}, \citenamefont
  {Kampert}, \citenamefont {Katz}, \citenamefont {Kawanai}, \citenamefont
  {Kovacs}, \citenamefont {Mages}, \citenamefont {Pasztor}, \citenamefont
  {Pittler}, \citenamefont {Redondo}, \citenamefont {Ringwald},\ and\
  \citenamefont {Szabo}}]{QCDCalI}%
  \BibitemOpen
  \bibfield  {author} {\bibinfo {author} {\bibfnamefont {S.}~\bibnamefont
  {Borsanyi}}, \bibinfo {author} {\bibfnamefont {Z.}~\bibnamefont {Fodor}},
  \bibinfo {author} {\bibfnamefont {J.}~\bibnamefont {Guenther}}, \bibinfo
  {author} {\bibfnamefont {K.-H.}\ \bibnamefont {Kampert}}, \bibinfo {author}
  {\bibfnamefont {S.~D.}\ \bibnamefont {Katz}}, \bibinfo {author}
  {\bibfnamefont {T.}~\bibnamefont {Kawanai}}, \bibinfo {author} {\bibfnamefont
  {T.~G.}\ \bibnamefont {Kovacs}}, \bibinfo {author} {\bibfnamefont {S.~W.}\
  \bibnamefont {Mages}}, \bibinfo {author} {\bibfnamefont {A.}~\bibnamefont
  {Pasztor}}, \bibinfo {author} {\bibfnamefont {F.}~\bibnamefont {Pittler}},
  \bibinfo {author} {\bibfnamefont {J.}~\bibnamefont {Redondo}}, \bibinfo
  {author} {\bibfnamefont {A.}~\bibnamefont {Ringwald}},\ and\ \bibinfo
  {author} {\bibfnamefont {K.~K.}\ \bibnamefont {Szabo}},\ }\href
  {https://doi.org/10.1038/nature20115} {\bibfield  {journal} {\bibinfo
  {journal} {Nature}\ }\textbf {\bibinfo {volume} {539}},\ \bibinfo {pages}
  {69} (\bibinfo {year} {2016})}\BibitemShut {NoStop}%
\bibitem [{\citenamefont {Dine}\ \emph {et~al.}(2017)\citenamefont {Dine},
  \citenamefont {Draper}, \citenamefont {Stephenson-Haskins},\ and\
  \citenamefont {Xu}}]{QCDCalII}%
  \BibitemOpen
  \bibfield  {author} {\bibinfo {author} {\bibfnamefont {M.}~\bibnamefont
  {Dine}}, \bibinfo {author} {\bibfnamefont {P.}~\bibnamefont {Draper}},
  \bibinfo {author} {\bibfnamefont {L.}~\bibnamefont {Stephenson-Haskins}},\
  and\ \bibinfo {author} {\bibfnamefont {D.}~\bibnamefont {Xu}},\ }\href
  {https://doi.org/10.1103/PhysRevD.96.095001} {\bibfield  {journal} {\bibinfo
  {journal} {Phys. Rev. D}\ }\textbf {\bibinfo {volume} {96}},\ \bibinfo
  {pages} {095001} (\bibinfo {year} {2017})}\BibitemShut {NoStop}%
\bibitem [{\citenamefont {Hiramatsu}\ \emph {et~al.}(2011)\citenamefont
  {Hiramatsu}, \citenamefont {Kawasaki}, \citenamefont {Sekiguchi},
  \citenamefont {Yamaguchi},\ and\ \citenamefont {Yokoyama}}]{QCDCalIII}%
  \BibitemOpen
  \bibfield  {author} {\bibinfo {author} {\bibfnamefont {T.}~\bibnamefont
  {Hiramatsu}}, \bibinfo {author} {\bibfnamefont {M.}~\bibnamefont {Kawasaki}},
  \bibinfo {author} {\bibfnamefont {T.}~\bibnamefont {Sekiguchi}}, \bibinfo
  {author} {\bibfnamefont {M.}~\bibnamefont {Yamaguchi}},\ and\ \bibinfo
  {author} {\bibfnamefont {J.}~\bibnamefont {Yokoyama}},\ }\href
  {https://doi.org/10.1103/PhysRevD.83.123531} {\bibfield  {journal} {\bibinfo
  {journal} {Phys. Rev. D}\ }\textbf {\bibinfo {volume} {83}},\ \bibinfo
  {pages} {123531} (\bibinfo {year} {2011})}\BibitemShut {NoStop}%
\bibitem [{\citenamefont {Kawasaki}\ \emph {et~al.}(2015)\citenamefont
  {Kawasaki}, \citenamefont {Saikawa},\ and\ \citenamefont
  {Sekiguchi}}]{QCDCalIV}%
  \BibitemOpen
  \bibfield  {author} {\bibinfo {author} {\bibfnamefont {M.}~\bibnamefont
  {Kawasaki}}, \bibinfo {author} {\bibfnamefont {K.}~\bibnamefont {Saikawa}},\
  and\ \bibinfo {author} {\bibfnamefont {T.}~\bibnamefont {Sekiguchi}},\ }\href
  {https://doi.org/10.1103/PhysRevD.91.065014} {\bibfield  {journal} {\bibinfo
  {journal} {Phys. Rev. D}\ }\textbf {\bibinfo {volume} {91}},\ \bibinfo
  {pages} {065014} (\bibinfo {year} {2015})}\BibitemShut {NoStop}%
\bibitem [{\citenamefont {Berkowitz}\ \emph {et~al.}(2015)\citenamefont
  {Berkowitz}, \citenamefont {Buchoff},\ and\ \citenamefont
  {Rinaldi}}]{QCDCalV}%
  \BibitemOpen
  \bibfield  {author} {\bibinfo {author} {\bibfnamefont {E.}~\bibnamefont
  {Berkowitz}}, \bibinfo {author} {\bibfnamefont {M.~I.}\ \bibnamefont
  {Buchoff}},\ and\ \bibinfo {author} {\bibfnamefont {E.}~\bibnamefont
  {Rinaldi}},\ }\href {https://doi.org/10.1103/PhysRevD.92.034507} {\bibfield
  {journal} {\bibinfo  {journal} {Phys. Rev. D}\ }\textbf {\bibinfo {volume}
  {92}},\ \bibinfo {pages} {034507} (\bibinfo {year} {2015})}\BibitemShut
  {NoStop}%
\bibitem [{\citenamefont {Fleury}\ and\ \citenamefont
  {Moore}(2016)}]{QCDCalVI}%
  \BibitemOpen
  \bibfield  {author} {\bibinfo {author} {\bibfnamefont {L.}~\bibnamefont
  {Fleury}}\ and\ \bibinfo {author} {\bibfnamefont {G.~D.}\ \bibnamefont
  {Moore}},\ }\href {https://doi.org/10.1088/1475-7516/2016/01/004} {\bibfield
  {journal} {\bibinfo  {journal} {J. Cosmol. Astropart. Phys.}\ }\textbf
  {\bibinfo {volume} {01}}\bibinfo  {number} { (2016)},\ \bibinfo {pages}
  {004}}\BibitemShut {NoStop}%
\bibitem [{\citenamefont {Bonati}\ \emph {et~al.}(2016)\citenamefont {Bonati},
  \citenamefont {D'Elia}, \citenamefont {Mariti}, \citenamefont {Martinelli},
  \citenamefont {Mesiti}, \citenamefont {Negro}, \citenamefont {Sanfilippo},\
  and\ \citenamefont {Villadoro}}]{QCDCalVII}%
  \BibitemOpen
\bibfield  {number} {  }\bibfield  {author} {\bibinfo {author} {\bibfnamefont
  {C.}~\bibnamefont {Bonati}}, \bibinfo {author} {\bibfnamefont
  {M.}~\bibnamefont {D'Elia}}, \bibinfo {author} {\bibfnamefont
  {M.}~\bibnamefont {Mariti}}, \bibinfo {author} {\bibfnamefont
  {G.}~\bibnamefont {Martinelli}}, \bibinfo {author} {\bibfnamefont
  {M.}~\bibnamefont {Mesiti}}, \bibinfo {author} {\bibfnamefont
  {F.}~\bibnamefont {Negro}}, \bibinfo {author} {\bibfnamefont
  {F.}~\bibnamefont {Sanfilippo}},\ and\ \bibinfo {author} {\bibfnamefont
  {G.}~\bibnamefont {Villadoro}},\ }\href
  {https://doi.org/10.1007/JHEP03(2016)155} {\bibfield  {journal} {\bibinfo
  {journal} {JHEP}\ }\textbf {\bibinfo {volume} {03}}\bibinfo  {number} {
  (2016)},\ \bibinfo {pages} {155}}\BibitemShut {NoStop}%
\bibitem [{\citenamefont {Petreczky}\ \emph {et~al.}(2016)\citenamefont
  {Petreczky}, \citenamefont {Schadler},\ and\ \citenamefont
  {Sharma}}]{QCDCalVIII}%
  \BibitemOpen
\bibfield  {number} {  }\bibfield  {author} {\bibinfo {author} {\bibfnamefont
  {P.}~\bibnamefont {Petreczky}}, \bibinfo {author} {\bibfnamefont {H.-P.}\
  \bibnamefont {Schadler}},\ and\ \bibinfo {author} {\bibfnamefont
  {S.}~\bibnamefont {Sharma}},\ }\href
  {https://doi.org/10.1016/j.physletb.2016.09.063} {\bibfield  {journal}
  {\bibinfo  {journal} {Phys. Lett. B}\ }\textbf {\bibinfo {volume} {762}},\
  \bibinfo {pages} {498} (\bibinfo {year} {2016})}\BibitemShut {NoStop}%
\bibitem [{\citenamefont {Ballesteros}\ \emph {et~al.}(2017)\citenamefont
  {Ballesteros}, \citenamefont {Redondo}, \citenamefont {Ringwald},\ and\
  \citenamefont {Tamarit}}]{QCDCalIX}%
  \BibitemOpen
  \bibfield  {author} {\bibinfo {author} {\bibfnamefont {G.}~\bibnamefont
  {Ballesteros}}, \bibinfo {author} {\bibfnamefont {J.}~\bibnamefont
  {Redondo}}, \bibinfo {author} {\bibfnamefont {A.}~\bibnamefont {Ringwald}},\
  and\ \bibinfo {author} {\bibfnamefont {C.}~\bibnamefont {Tamarit}},\ }\href
  {https://doi.org/10.1103/PhysRevLett.118.071802} {\bibfield  {journal}
  {\bibinfo  {journal} {Phys. Rev. Lett.}\ }\textbf {\bibinfo {volume} {118}},\
  \bibinfo {pages} {071802} (\bibinfo {year} {2017})}\BibitemShut {NoStop}%
\bibitem [{\citenamefont {Klaer}\ and\ \citenamefont {Moore}(2017)}]{QCDCalX}%
  \BibitemOpen
  \bibfield  {author} {\bibinfo {author} {\bibfnamefont {V.~B.}\ \bibnamefont
  {Klaer}}\ and\ \bibinfo {author} {\bibfnamefont {G.~D.}\ \bibnamefont
  {Moore}},\ }\href {https://doi.org/10.1088/1475-7516/2017/11/049} {\bibfield
  {journal} {\bibinfo  {journal} {J. Cosmol. Astropart. Phys.}\ }\textbf
  {\bibinfo {volume} {11}}\bibinfo  {number} { (2017)},\ \bibinfo {pages}
  {049}}\BibitemShut {NoStop}%
\bibitem [{\citenamefont {Buschmann}\ \emph {et~al.}(2020)\citenamefont
  {Buschmann}, \citenamefont {Foster},\ and\ \citenamefont {Safdi}}]{QCDCalXI}%
  \BibitemOpen
\bibfield  {number} {  }\bibfield  {author} {\bibinfo {author} {\bibfnamefont
  {M.}~\bibnamefont {Buschmann}}, \bibinfo {author} {\bibfnamefont {J.~W.}\
  \bibnamefont {Foster}},\ and\ \bibinfo {author} {\bibfnamefont {B.~R.}\
  \bibnamefont {Safdi}},\ }\href
  {https://doi.org/10.1103/PhysRevLett.124.161103} {\bibfield  {journal}
  {\bibinfo  {journal} {Phys. Rev. Lett.}\ }\textbf {\bibinfo {volume} {124}},\
  \bibinfo {pages} {161103} (\bibinfo {year} {2020})}\BibitemShut {NoStop}%
\bibitem [{\citenamefont {Gorghetto}\ \emph {et~al.}(2021)\citenamefont
  {Gorghetto}, \citenamefont {Hardy},\ and\ \citenamefont
  {Villadoro}}]{QCDCalXII}%
  \BibitemOpen
  \bibfield  {author} {\bibinfo {author} {\bibfnamefont {M.}~\bibnamefont
  {Gorghetto}}, \bibinfo {author} {\bibfnamefont {E.}~\bibnamefont {Hardy}},\
  and\ \bibinfo {author} {\bibfnamefont {G.}~\bibnamefont {Villadoro}},\ }\href
  {https://doi.org/10.21468/SciPostPhys.10.2.050} {\bibfield  {journal}
  {\bibinfo  {journal} {SciPost Phys.}\ }\textbf {\bibinfo {volume} {10}},\
  \bibinfo {pages} {050} (\bibinfo {year} {2021})}\BibitemShut {NoStop}%
\bibitem [{\citenamefont {Buschmann}\ \emph {et~al.}(2022)\citenamefont
  {Buschmann}, \citenamefont {Foster}, \citenamefont {Hook}, \citenamefont
  {Peterson}, \citenamefont {Willcox}, \citenamefont {Zhang},\ and\
  \citenamefont {Safdi}}]{QCDCalXIII}%
  \BibitemOpen
  \bibfield  {author} {\bibinfo {author} {\bibfnamefont {M.}~\bibnamefont
  {Buschmann}}, \bibinfo {author} {\bibfnamefont {J.~W.}\ \bibnamefont
  {Foster}}, \bibinfo {author} {\bibfnamefont {A.}~\bibnamefont {Hook}},
  \bibinfo {author} {\bibfnamefont {A.}~\bibnamefont {Peterson}}, \bibinfo
  {author} {\bibfnamefont {D.~E.}\ \bibnamefont {Willcox}}, \bibinfo {author}
  {\bibfnamefont {W.}~\bibnamefont {Zhang}},\ and\ \bibinfo {author}
  {\bibfnamefont {B.~R.}\ \bibnamefont {Safdi}},\ }\href
  {https://doi.org/10.1038/s41467-022-28669-y} {\bibfield  {journal} {\bibinfo
  {journal} {Nature Commun.}\ }\textbf {\bibinfo {volume} {13}},\ \bibinfo
  {pages} {1049} (\bibinfo {year} {2022})}\BibitemShut {NoStop}%
\bibitem [{\citenamefont {Sikivie}(1983)}]{SikivieI}%
  \BibitemOpen
  \bibfield  {author} {\bibinfo {author} {\bibfnamefont {P.}~\bibnamefont
  {Sikivie}},\ }\href {https://doi.org/10.1103/PhysRevLett.51.1415} {\bibfield
  {journal} {\bibinfo  {journal} {Phys. Rev. Lett.}\ }\textbf {\bibinfo
  {volume} {51}},\ \bibinfo {pages} {1415} (\bibinfo {year}
  {1983})}\BibitemShut {NoStop}%
\bibitem [{\citenamefont {Sikivie}(1985)}]{SikivieII}%
  \BibitemOpen
  \bibfield  {author} {\bibinfo {author} {\bibfnamefont {P.}~\bibnamefont
  {Sikivie}},\ }\href {https://doi.org/10.1103/PhysRevD.32.2988} {\bibfield
  {journal} {\bibinfo  {journal} {Phys. Rev. D}\ }\textbf {\bibinfo {volume}
  {32}},\ \bibinfo {pages} {2988} (\bibinfo {year} {1985})}\BibitemShut
  {NoStop}%
\bibitem [{\citenamefont {Hagmann}\ \emph {et~al.}(1998)\citenamefont
  {Hagmann}, \citenamefont {Kinion}, \citenamefont {Stoeffl}, \citenamefont
  {van Bibber}, \citenamefont {Daw}, \citenamefont {Peng}, \citenamefont
  {Rosenberg}, \citenamefont {LaVeigne}, \citenamefont {Sikivie}, \citenamefont
  {Sullivan}, \citenamefont {Tanner}, \citenamefont {Nezrick}, \citenamefont
  {Turner}, \citenamefont {Moltz}, \citenamefont {Powell},\ and\ \citenamefont
  {Golubev}}]{ADMXI}%
  \BibitemOpen
  \bibfield  {author} {\bibinfo {author} {\bibfnamefont {C.}~\bibnamefont
  {Hagmann}}, \bibinfo {author} {\bibfnamefont {D.}~\bibnamefont {Kinion}},
  \bibinfo {author} {\bibfnamefont {W.}~\bibnamefont {Stoeffl}}, \bibinfo
  {author} {\bibfnamefont {K.}~\bibnamefont {van Bibber}}, \bibinfo {author}
  {\bibfnamefont {E.}~\bibnamefont {Daw}}, \bibinfo {author} {\bibfnamefont
  {H.}~\bibnamefont {Peng}}, \bibinfo {author} {\bibfnamefont {L.~J.}\
  \bibnamefont {Rosenberg}}, \bibinfo {author} {\bibfnamefont {J.}~\bibnamefont
  {LaVeigne}}, \bibinfo {author} {\bibfnamefont {P.}~\bibnamefont {Sikivie}},
  \bibinfo {author} {\bibfnamefont {N.~S.}\ \bibnamefont {Sullivan}}, \bibinfo
  {author} {\bibfnamefont {D.~B.}\ \bibnamefont {Tanner}}, \bibinfo {author}
  {\bibfnamefont {F.}~\bibnamefont {Nezrick}}, \bibinfo {author} {\bibfnamefont
  {M.~S.}\ \bibnamefont {Turner}}, \bibinfo {author} {\bibfnamefont {D.~M.}\
  \bibnamefont {Moltz}}, \bibinfo {author} {\bibfnamefont {J.}~\bibnamefont
  {Powell}},\ and\ \bibinfo {author} {\bibfnamefont {N.~A.}\ \bibnamefont
  {Golubev}},\ }\href {https://doi.org/10.1103/PhysRevLett.80.2043} {\bibfield
  {journal} {\bibinfo  {journal} {Phys. Rev. Lett.}\ }\textbf {\bibinfo
  {volume} {80}},\ \bibinfo {pages} {2043} (\bibinfo {year}
  {1998})}\BibitemShut {NoStop}%
\bibitem [{\citenamefont {Asztalos}\ \emph {et~al.}(2002)\citenamefont
  {Asztalos}, \citenamefont {Daw}, \citenamefont {Peng}, \citenamefont
  {Rosenberg}, \citenamefont {Yu}, \citenamefont {Hagmann}, \citenamefont
  {Kinion}, \citenamefont {Stoeffl}, \citenamefont {van Bibber}, \citenamefont
  {LaVeigne}, \citenamefont {Sikivie}, \citenamefont {Sullivan}, \citenamefont
  {Tanner}, \citenamefont {Nezrick},\ and\ \citenamefont {Moltz}}]{ADMXII}%
  \BibitemOpen
  \bibfield  {author} {\bibinfo {author} {\bibfnamefont {S.~J.}\ \bibnamefont
  {Asztalos}}, \bibinfo {author} {\bibfnamefont {E.}~\bibnamefont {Daw}},
  \bibinfo {author} {\bibfnamefont {H.}~\bibnamefont {Peng}}, \bibinfo {author}
  {\bibfnamefont {L.~J.}\ \bibnamefont {Rosenberg}}, \bibinfo {author}
  {\bibfnamefont {D.~B.}\ \bibnamefont {Yu}}, \bibinfo {author} {\bibfnamefont
  {C.}~\bibnamefont {Hagmann}}, \bibinfo {author} {\bibfnamefont
  {D.}~\bibnamefont {Kinion}}, \bibinfo {author} {\bibfnamefont
  {W.}~\bibnamefont {Stoeffl}}, \bibinfo {author} {\bibfnamefont
  {K.}~\bibnamefont {van Bibber}}, \bibinfo {author} {\bibfnamefont
  {J.}~\bibnamefont {LaVeigne}}, \bibinfo {author} {\bibfnamefont
  {P.}~\bibnamefont {Sikivie}}, \bibinfo {author} {\bibfnamefont {N.~S.}\
  \bibnamefont {Sullivan}}, \bibinfo {author} {\bibfnamefont {D.~B.}\
  \bibnamefont {Tanner}}, \bibinfo {author} {\bibfnamefont {F.}~\bibnamefont
  {Nezrick}},\ and\ \bibinfo {author} {\bibfnamefont {D.~M.}\ \bibnamefont
  {Moltz}},\ }\href {https://doi.org/10.1086/341130} {\bibfield  {journal}
  {\bibinfo  {journal} {The Astrophysical Journal}\ }\textbf {\bibinfo {volume}
  {571}},\ \bibinfo {pages} {L27} (\bibinfo {year} {2002})}\BibitemShut
  {NoStop}%
\bibitem [{\citenamefont {Asztalos}\ \emph {et~al.}(2004)\citenamefont
  {Asztalos}, \citenamefont {Bradley}, \citenamefont {Duffy}, \citenamefont
  {Hagmann}, \citenamefont {Kinion}, \citenamefont {Moltz}, \citenamefont
  {Rosenberg}, \citenamefont {Sikivie}, \citenamefont {Stoeffl}, \citenamefont
  {Sullivan}, \citenamefont {Tanner}, \citenamefont {van Bibber},\ and\
  \citenamefont {Yu}}]{ADMXIII}%
  \BibitemOpen
  \bibfield  {author} {\bibinfo {author} {\bibfnamefont {S.~J.}\ \bibnamefont
  {Asztalos}}, \bibinfo {author} {\bibfnamefont {R.~F.}\ \bibnamefont
  {Bradley}}, \bibinfo {author} {\bibfnamefont {L.}~\bibnamefont {Duffy}},
  \bibinfo {author} {\bibfnamefont {C.}~\bibnamefont {Hagmann}}, \bibinfo
  {author} {\bibfnamefont {D.}~\bibnamefont {Kinion}}, \bibinfo {author}
  {\bibfnamefont {D.~M.}\ \bibnamefont {Moltz}}, \bibinfo {author}
  {\bibfnamefont {L.~J.}\ \bibnamefont {Rosenberg}}, \bibinfo {author}
  {\bibfnamefont {P.}~\bibnamefont {Sikivie}}, \bibinfo {author} {\bibfnamefont
  {W.}~\bibnamefont {Stoeffl}}, \bibinfo {author} {\bibfnamefont {N.~S.}\
  \bibnamefont {Sullivan}}, \bibinfo {author} {\bibfnamefont {D.~B.}\
  \bibnamefont {Tanner}}, \bibinfo {author} {\bibfnamefont {K.}~\bibnamefont
  {van Bibber}},\ and\ \bibinfo {author} {\bibfnamefont {D.~B.}\ \bibnamefont
  {Yu}},\ }\href {https://doi.org/10.1103/PhysRevD.69.011101} {\bibfield
  {journal} {\bibinfo  {journal} {Phys. Rev. D}\ }\textbf {\bibinfo {volume}
  {69}},\ \bibinfo {pages} {011101(R)} (\bibinfo {year} {2004})}\BibitemShut
  {NoStop}%
\bibitem [{\citenamefont {Asztalos}\ \emph {et~al.}(2010)\citenamefont
  {Asztalos}, \citenamefont {Carosi}, \citenamefont {Hagmann}, \citenamefont
  {Kinion}, \citenamefont {van Bibber}, \citenamefont {Hotz}, \citenamefont
  {Rosenberg}, \citenamefont {Rybka}, \citenamefont {Hoskins}, \citenamefont
  {Hwang}, \citenamefont {Sikivie}, \citenamefont {Tanner}, \citenamefont
  {Bradley},\ and\ \citenamefont {Clarke}}]{ADMXIV}%
  \BibitemOpen
  \bibfield  {author} {\bibinfo {author} {\bibfnamefont {S.~J.}\ \bibnamefont
  {Asztalos}}, \bibinfo {author} {\bibfnamefont {G.}~\bibnamefont {Carosi}},
  \bibinfo {author} {\bibfnamefont {C.}~\bibnamefont {Hagmann}}, \bibinfo
  {author} {\bibfnamefont {D.}~\bibnamefont {Kinion}}, \bibinfo {author}
  {\bibfnamefont {K.}~\bibnamefont {van Bibber}}, \bibinfo {author}
  {\bibfnamefont {M.}~\bibnamefont {Hotz}}, \bibinfo {author} {\bibfnamefont
  {L.~J.}\ \bibnamefont {Rosenberg}}, \bibinfo {author} {\bibfnamefont
  {G.}~\bibnamefont {Rybka}}, \bibinfo {author} {\bibfnamefont
  {J.}~\bibnamefont {Hoskins}}, \bibinfo {author} {\bibfnamefont
  {J.}~\bibnamefont {Hwang}}, \bibinfo {author} {\bibfnamefont
  {P.}~\bibnamefont {Sikivie}}, \bibinfo {author} {\bibfnamefont {D.~B.}\
  \bibnamefont {Tanner}}, \bibinfo {author} {\bibfnamefont {R.}~\bibnamefont
  {Bradley}},\ and\ \bibinfo {author} {\bibfnamefont {J.}~\bibnamefont
  {Clarke}},\ }\href {https://doi.org/10.1103/PhysRevLett.104.041301}
  {\bibfield  {journal} {\bibinfo  {journal} {Phys. Rev. Lett.}\ }\textbf
  {\bibinfo {volume} {104}},\ \bibinfo {pages} {041301} (\bibinfo {year}
  {2010})}\BibitemShut {NoStop}%
\bibitem [{\citenamefont {Du}\ \emph {et~al.}(2018)\citenamefont {Du},
  \citenamefont {Force}, \citenamefont {Khatiwada}, \citenamefont {Lentz},
  \citenamefont {Ottens}, \citenamefont {Rosenberg}, \citenamefont {Rybka},
  \citenamefont {Carosi}, \citenamefont {Woollett}, \citenamefont {Bowring},
  \citenamefont {Chou}, \citenamefont {Sonnenschein}, \citenamefont {Wester},
  \citenamefont {Boutan}, \citenamefont {Oblath}, \citenamefont {Bradley},
  \citenamefont {Daw}, \citenamefont {Dixit}, \citenamefont {Clarke},
  \citenamefont {O'Kelley}, \citenamefont {Crisosto}, \citenamefont {Gleason},
  \citenamefont {Jois}, \citenamefont {Sikivie}, \citenamefont {Stern},
  \citenamefont {Sullivan}, \citenamefont {Tanner},\ and\ \citenamefont
  {Hilton}}]{ADMXV}%
  \BibitemOpen
  \bibfield  {author} {\bibinfo {author} {\bibfnamefont {N.}~\bibnamefont
  {Du}}, \bibinfo {author} {\bibfnamefont {N.}~\bibnamefont {Force}}, \bibinfo
  {author} {\bibfnamefont {R.}~\bibnamefont {Khatiwada}}, \bibinfo {author}
  {\bibfnamefont {E.}~\bibnamefont {Lentz}}, \bibinfo {author} {\bibfnamefont
  {R.}~\bibnamefont {Ottens}}, \bibinfo {author} {\bibfnamefont {L.~J.}\
  \bibnamefont {Rosenberg}}, \bibinfo {author} {\bibfnamefont {G.}~\bibnamefont
  {Rybka}}, \bibinfo {author} {\bibfnamefont {G.}~\bibnamefont {Carosi}},
  \bibinfo {author} {\bibfnamefont {N.}~\bibnamefont {Woollett}}, \bibinfo
  {author} {\bibfnamefont {D.}~\bibnamefont {Bowring}}, \bibinfo {author}
  {\bibfnamefont {A.~S.}\ \bibnamefont {Chou}}, \bibinfo {author}
  {\bibfnamefont {A.}~\bibnamefont {Sonnenschein}}, \bibinfo {author}
  {\bibfnamefont {W.}~\bibnamefont {Wester}}, \bibinfo {author} {\bibfnamefont
  {C.}~\bibnamefont {Boutan}}, \bibinfo {author} {\bibfnamefont {N.~S.}\
  \bibnamefont {Oblath}}, \bibinfo {author} {\bibfnamefont {R.}~\bibnamefont
  {Bradley}}, \bibinfo {author} {\bibfnamefont {E.~J.}\ \bibnamefont {Daw}},
  \bibinfo {author} {\bibfnamefont {A.~V.}\ \bibnamefont {Dixit}}, \bibinfo
  {author} {\bibfnamefont {J.}~\bibnamefont {Clarke}}, \bibinfo {author}
  {\bibfnamefont {S.~R.}\ \bibnamefont {O'Kelley}}, \bibinfo {author}
  {\bibfnamefont {N.}~\bibnamefont {Crisosto}}, \bibinfo {author}
  {\bibfnamefont {J.~R.}\ \bibnamefont {Gleason}}, \bibinfo {author}
  {\bibfnamefont {S.}~\bibnamefont {Jois}}, \bibinfo {author} {\bibfnamefont
  {P.}~\bibnamefont {Sikivie}}, \bibinfo {author} {\bibfnamefont
  {I.}~\bibnamefont {Stern}}, \bibinfo {author} {\bibfnamefont {N.~S.}\
  \bibnamefont {Sullivan}}, \bibinfo {author} {\bibfnamefont {D.~B.}\
  \bibnamefont {Tanner}},\ and\ \bibinfo {author} {\bibfnamefont {G.~C.}\
  \bibnamefont {Hilton}} (\bibinfo {collaboration} {ADMX Collaboration}),\
  }\href {https://doi.org/10.1103/PhysRevLett.120.151301} {\bibfield  {journal}
  {\bibinfo  {journal} {Phys. Rev. Lett.}\ }\textbf {\bibinfo {volume} {120}},\
  \bibinfo {pages} {151301} (\bibinfo {year} {2018})}\BibitemShut {NoStop}%
\bibitem [{\citenamefont {Braine}\ \emph {et~al.}(2020)\citenamefont {Braine},
  \citenamefont {Cervantes}, \citenamefont {Crisosto}, \citenamefont {Du},
  \citenamefont {Kimes}, \citenamefont {Rosenberg}, \citenamefont {Rybka},
  \citenamefont {Yang}, \citenamefont {Bowring}, \citenamefont {Chou},
  \citenamefont {Khatiwada}, \citenamefont {Sonnenschein}, \citenamefont
  {Wester}, \citenamefont {Carosi}, \citenamefont {Woollett}, \citenamefont
  {Duffy}, \citenamefont {Bradley}, \citenamefont {Boutan}, \citenamefont
  {Jones}, \citenamefont {LaRoque}, \citenamefont {Oblath}, \citenamefont
  {Taubman}, \citenamefont {Clarke}, \citenamefont {Dove}, \citenamefont
  {Eddins}, \citenamefont {O'Kelley}, \citenamefont {Nawaz}, \citenamefont
  {Siddiqi}, \citenamefont {Stevenson}, \citenamefont {Agrawal}, \citenamefont
  {Dixit}, \citenamefont {Gleason}, \citenamefont {Jois}, \citenamefont
  {Sikivie}, \citenamefont {Solomon}, \citenamefont {Sullivan}, \citenamefont
  {Tanner}, \citenamefont {Lentz}, \citenamefont {Daw}, \citenamefont
  {Buckley}, \citenamefont {Harrington}, \citenamefont {Henriksen},\ and\
  \citenamefont {Murch}}]{ADMXVI}%
  \BibitemOpen
  \bibfield  {author} {\bibinfo {author} {\bibfnamefont {T.}~\bibnamefont
  {Braine}}, \bibinfo {author} {\bibfnamefont {R.}~\bibnamefont {Cervantes}},
  \bibinfo {author} {\bibfnamefont {N.}~\bibnamefont {Crisosto}}, \bibinfo
  {author} {\bibfnamefont {N.}~\bibnamefont {Du}}, \bibinfo {author}
  {\bibfnamefont {S.}~\bibnamefont {Kimes}}, \bibinfo {author} {\bibfnamefont
  {L.~J.}\ \bibnamefont {Rosenberg}}, \bibinfo {author} {\bibfnamefont
  {G.}~\bibnamefont {Rybka}}, \bibinfo {author} {\bibfnamefont
  {J.}~\bibnamefont {Yang}}, \bibinfo {author} {\bibfnamefont {D.}~\bibnamefont
  {Bowring}}, \bibinfo {author} {\bibfnamefont {A.~S.}\ \bibnamefont {Chou}},
  \bibinfo {author} {\bibfnamefont {R.}~\bibnamefont {Khatiwada}}, \bibinfo
  {author} {\bibfnamefont {A.}~\bibnamefont {Sonnenschein}}, \bibinfo {author}
  {\bibfnamefont {W.}~\bibnamefont {Wester}}, \bibinfo {author} {\bibfnamefont
  {G.}~\bibnamefont {Carosi}}, \bibinfo {author} {\bibfnamefont
  {N.}~\bibnamefont {Woollett}}, \bibinfo {author} {\bibfnamefont {L.~D.}\
  \bibnamefont {Duffy}}, \bibinfo {author} {\bibfnamefont {R.}~\bibnamefont
  {Bradley}}, \bibinfo {author} {\bibfnamefont {C.}~\bibnamefont {Boutan}},
  \bibinfo {author} {\bibfnamefont {M.}~\bibnamefont {Jones}}, \bibinfo
  {author} {\bibfnamefont {B.~H.}\ \bibnamefont {LaRoque}}, \bibinfo {author}
  {\bibfnamefont {N.~S.}\ \bibnamefont {Oblath}}, \bibinfo {author}
  {\bibfnamefont {M.~S.}\ \bibnamefont {Taubman}}, \bibinfo {author}
  {\bibfnamefont {J.}~\bibnamefont {Clarke}}, \bibinfo {author} {\bibfnamefont
  {A.}~\bibnamefont {Dove}}, \bibinfo {author} {\bibfnamefont {A.}~\bibnamefont
  {Eddins}}, \bibinfo {author} {\bibfnamefont {S.~R.}\ \bibnamefont
  {O'Kelley}}, \bibinfo {author} {\bibfnamefont {S.}~\bibnamefont {Nawaz}},
  \bibinfo {author} {\bibfnamefont {I.}~\bibnamefont {Siddiqi}}, \bibinfo
  {author} {\bibfnamefont {N.}~\bibnamefont {Stevenson}}, \bibinfo {author}
  {\bibfnamefont {A.}~\bibnamefont {Agrawal}}, \bibinfo {author} {\bibfnamefont
  {A.~V.}\ \bibnamefont {Dixit}}, \bibinfo {author} {\bibfnamefont {J.~R.}\
  \bibnamefont {Gleason}}, \bibinfo {author} {\bibfnamefont {S.}~\bibnamefont
  {Jois}}, \bibinfo {author} {\bibfnamefont {P.}~\bibnamefont {Sikivie}},
  \bibinfo {author} {\bibfnamefont {J.~A.}\ \bibnamefont {Solomon}}, \bibinfo
  {author} {\bibfnamefont {N.~S.}\ \bibnamefont {Sullivan}}, \bibinfo {author}
  {\bibfnamefont {D.~B.}\ \bibnamefont {Tanner}}, \bibinfo {author}
  {\bibfnamefont {E.}~\bibnamefont {Lentz}}, \bibinfo {author} {\bibfnamefont
  {E.~J.}\ \bibnamefont {Daw}}, \bibinfo {author} {\bibfnamefont {J.~H.}\
  \bibnamefont {Buckley}}, \bibinfo {author} {\bibfnamefont {P.~M.}\
  \bibnamefont {Harrington}}, \bibinfo {author} {\bibfnamefont {E.~A.}\
  \bibnamefont {Henriksen}},\ and\ \bibinfo {author} {\bibfnamefont {K.~W.}\
  \bibnamefont {Murch}} (\bibinfo {collaboration} {ADMX Collaboration}),\
  }\href {https://doi.org/10.1103/PhysRevLett.124.101303} {\bibfield  {journal}
  {\bibinfo  {journal} {Phys. Rev. Lett.}\ }\textbf {\bibinfo {volume} {124}},\
  \bibinfo {pages} {101303} (\bibinfo {year} {2020})}\BibitemShut {NoStop}%
\bibitem [{\citenamefont {Bartram}\ \emph
  {et~al.}(2021{\natexlab{a}})\citenamefont {Bartram} \emph
  {et~al.}}]{ADMXVII}%
  \BibitemOpen
  \bibfield  {author} {\bibinfo {author} {\bibfnamefont {C.}~\bibnamefont
  {Bartram}} \emph {et~al.} (\bibinfo {collaboration} {ADMX Collaboration}),\
  }\href {https://doi.org/10.1103/PhysRevLett.127.261803} {\bibfield  {journal}
  {\bibinfo  {journal} {Phys. Rev. Lett.}\ }\textbf {\bibinfo {volume} {127}},\
  \bibinfo {pages} {261803} (\bibinfo {year} {2021}{\natexlab{a}})}\BibitemShut
  {NoStop}%
\bibitem [{\citenamefont {Brubaker}\ \emph
  {et~al.}(2017{\natexlab{a}})\citenamefont {Brubaker}, \citenamefont {Zhong},
  \citenamefont {Gurevich}, \citenamefont {Cahn}, \citenamefont {Lamoreaux},
  \citenamefont {Simanovskaia}, \citenamefont {Root}, \citenamefont {Lewis},
  \citenamefont {Al~Kenany}, \citenamefont {Backes}, \citenamefont {Urdinaran},
  \citenamefont {Rapidis}, \citenamefont {Shokair}, \citenamefont {van Bibber},
  \citenamefont {Palken}, \citenamefont {Malnou}, \citenamefont {Kindel},
  \citenamefont {Anil}, \citenamefont {Lehnert},\ and\ \citenamefont
  {Carosi}}]{HAYSTACIII}%
  \BibitemOpen
  \bibfield  {author} {\bibinfo {author} {\bibfnamefont {B.~M.}\ \bibnamefont
  {Brubaker}}, \bibinfo {author} {\bibfnamefont {L.}~\bibnamefont {Zhong}},
  \bibinfo {author} {\bibfnamefont {Y.~V.}\ \bibnamefont {Gurevich}}, \bibinfo
  {author} {\bibfnamefont {S.~B.}\ \bibnamefont {Cahn}}, \bibinfo {author}
  {\bibfnamefont {S.~K.}\ \bibnamefont {Lamoreaux}}, \bibinfo {author}
  {\bibfnamefont {M.}~\bibnamefont {Simanovskaia}}, \bibinfo {author}
  {\bibfnamefont {J.~R.}\ \bibnamefont {Root}}, \bibinfo {author}
  {\bibfnamefont {S.~M.}\ \bibnamefont {Lewis}}, \bibinfo {author}
  {\bibfnamefont {S.}~\bibnamefont {Al~Kenany}}, \bibinfo {author}
  {\bibfnamefont {K.~M.}\ \bibnamefont {Backes}}, \bibinfo {author}
  {\bibfnamefont {I.}~\bibnamefont {Urdinaran}}, \bibinfo {author}
  {\bibfnamefont {N.~M.}\ \bibnamefont {Rapidis}}, \bibinfo {author}
  {\bibfnamefont {T.~M.}\ \bibnamefont {Shokair}}, \bibinfo {author}
  {\bibfnamefont {K.~A.}\ \bibnamefont {van Bibber}}, \bibinfo {author}
  {\bibfnamefont {D.~A.}\ \bibnamefont {Palken}}, \bibinfo {author}
  {\bibfnamefont {M.}~\bibnamefont {Malnou}}, \bibinfo {author} {\bibfnamefont
  {W.~F.}\ \bibnamefont {Kindel}}, \bibinfo {author} {\bibfnamefont {M.~A.}\
  \bibnamefont {Anil}}, \bibinfo {author} {\bibfnamefont {K.~W.}\ \bibnamefont
  {Lehnert}},\ and\ \bibinfo {author} {\bibfnamefont {G.}~\bibnamefont
  {Carosi}},\ }\href {https://doi.org/10.1103/PhysRevLett.118.061302}
  {\bibfield  {journal} {\bibinfo  {journal} {Phys. Rev. Lett.}\ }\textbf
  {\bibinfo {volume} {118}},\ \bibinfo {pages} {061302} (\bibinfo {year}
  {2017}{\natexlab{a}})}\BibitemShut {NoStop}%
\bibitem [{\citenamefont {Zhong}\ \emph {et~al.}(2018)\citenamefont {Zhong},
  \citenamefont {Al~Kenany}, \citenamefont {Backes}, \citenamefont {Brubaker},
  \citenamefont {Cahn}, \citenamefont {Carosi}, \citenamefont {Gurevich},
  \citenamefont {Kindel}, \citenamefont {Lamoreaux}, \citenamefont {Lehnert},
  \citenamefont {Lewis}, \citenamefont {Malnou}, \citenamefont {Maruyama},
  \citenamefont {Palken}, \citenamefont {Rapidis}, \citenamefont {Root},
  \citenamefont {Simanovskaia}, \citenamefont {Shokair}, \citenamefont
  {Speller}, \citenamefont {Urdinaran},\ and\ \citenamefont {van
  Bibber}}]{HAYSTACIV}%
  \BibitemOpen
  \bibfield  {author} {\bibinfo {author} {\bibfnamefont {L.}~\bibnamefont
  {Zhong}}, \bibinfo {author} {\bibfnamefont {S.}~\bibnamefont {Al~Kenany}},
  \bibinfo {author} {\bibfnamefont {K.~M.}\ \bibnamefont {Backes}}, \bibinfo
  {author} {\bibfnamefont {B.~M.}\ \bibnamefont {Brubaker}}, \bibinfo {author}
  {\bibfnamefont {S.~B.}\ \bibnamefont {Cahn}}, \bibinfo {author}
  {\bibfnamefont {G.}~\bibnamefont {Carosi}}, \bibinfo {author} {\bibfnamefont
  {Y.~V.}\ \bibnamefont {Gurevich}}, \bibinfo {author} {\bibfnamefont {W.~F.}\
  \bibnamefont {Kindel}}, \bibinfo {author} {\bibfnamefont {S.~K.}\
  \bibnamefont {Lamoreaux}}, \bibinfo {author} {\bibfnamefont {K.~W.}\
  \bibnamefont {Lehnert}}, \bibinfo {author} {\bibfnamefont {S.~M.}\
  \bibnamefont {Lewis}}, \bibinfo {author} {\bibfnamefont {M.}~\bibnamefont
  {Malnou}}, \bibinfo {author} {\bibfnamefont {R.~H.}\ \bibnamefont
  {Maruyama}}, \bibinfo {author} {\bibfnamefont {D.~A.}\ \bibnamefont
  {Palken}}, \bibinfo {author} {\bibfnamefont {N.~M.}\ \bibnamefont {Rapidis}},
  \bibinfo {author} {\bibfnamefont {J.~R.}\ \bibnamefont {Root}}, \bibinfo
  {author} {\bibfnamefont {M.}~\bibnamefont {Simanovskaia}}, \bibinfo {author}
  {\bibfnamefont {T.~M.}\ \bibnamefont {Shokair}}, \bibinfo {author}
  {\bibfnamefont {D.~H.}\ \bibnamefont {Speller}}, \bibinfo {author}
  {\bibfnamefont {I.}~\bibnamefont {Urdinaran}},\ and\ \bibinfo {author}
  {\bibfnamefont {K.~A.}\ \bibnamefont {van Bibber}},\ }\href
  {https://doi.org/10.1103/PhysRevD.97.092001} {\bibfield  {journal} {\bibinfo
  {journal} {Phys. Rev. D}\ }\textbf {\bibinfo {volume} {97}},\ \bibinfo
  {pages} {092001} (\bibinfo {year} {2018})}\BibitemShut {NoStop}%
\bibitem [{\citenamefont {Backes}\ \emph {et~al.}(2021)\citenamefont {Backes},
  \citenamefont {Palken}, \citenamefont {Al~Kenany}, \citenamefont {Brubaker},
  \citenamefont {Cahn}, \citenamefont {Droster}, \citenamefont {Hilton},
  \citenamefont {Ghosh}, \citenamefont {Jackson}, \citenamefont {Lamoreaux},
  \citenamefont {Leder}, \citenamefont {Lehnert}, \citenamefont {Lewis},
  \citenamefont {Malnou}, \citenamefont {Maruyama}, \citenamefont {Rapidis},
  \citenamefont {Simanovskaia}, \citenamefont {Singh}, \citenamefont {Speller},
  \citenamefont {Urdinaran}, \citenamefont {Vale}, \citenamefont {van
  Assendelft}, \citenamefont {van Bibber},\ and\ \citenamefont
  {Wang}}]{HAYSTACI}%
  \BibitemOpen
  \bibfield  {author} {\bibinfo {author} {\bibfnamefont {K.~M.}\ \bibnamefont
  {Backes}}, \bibinfo {author} {\bibfnamefont {D.~A.}\ \bibnamefont {Palken}},
  \bibinfo {author} {\bibfnamefont {S.}~\bibnamefont {Al~Kenany}}, \bibinfo
  {author} {\bibfnamefont {B.~M.}\ \bibnamefont {Brubaker}}, \bibinfo {author}
  {\bibfnamefont {S.~B.}\ \bibnamefont {Cahn}}, \bibinfo {author}
  {\bibfnamefont {A.}~\bibnamefont {Droster}}, \bibinfo {author} {\bibfnamefont
  {G.~C.}\ \bibnamefont {Hilton}}, \bibinfo {author} {\bibfnamefont
  {S.}~\bibnamefont {Ghosh}}, \bibinfo {author} {\bibfnamefont
  {H.}~\bibnamefont {Jackson}}, \bibinfo {author} {\bibfnamefont {S.~K.}\
  \bibnamefont {Lamoreaux}}, \bibinfo {author} {\bibfnamefont {A.~F.}\
  \bibnamefont {Leder}}, \bibinfo {author} {\bibfnamefont {K.~W.}\ \bibnamefont
  {Lehnert}}, \bibinfo {author} {\bibfnamefont {S.~M.}\ \bibnamefont {Lewis}},
  \bibinfo {author} {\bibfnamefont {M.}~\bibnamefont {Malnou}}, \bibinfo
  {author} {\bibfnamefont {R.~H.}\ \bibnamefont {Maruyama}}, \bibinfo {author}
  {\bibfnamefont {N.~M.}\ \bibnamefont {Rapidis}}, \bibinfo {author}
  {\bibfnamefont {M.}~\bibnamefont {Simanovskaia}}, \bibinfo {author}
  {\bibfnamefont {S.}~\bibnamefont {Singh}}, \bibinfo {author} {\bibfnamefont
  {D.~H.}\ \bibnamefont {Speller}}, \bibinfo {author} {\bibfnamefont
  {I.}~\bibnamefont {Urdinaran}}, \bibinfo {author} {\bibfnamefont {L.~R.}\
  \bibnamefont {Vale}}, \bibinfo {author} {\bibfnamefont {E.~C.}\ \bibnamefont
  {van Assendelft}}, \bibinfo {author} {\bibfnamefont {K.}~\bibnamefont {van
  Bibber}},\ and\ \bibinfo {author} {\bibfnamefont {H.}~\bibnamefont {Wang}},\
  }\href {https://doi.org/10.1038/s41586-021-03226-7} {\bibfield  {journal}
  {\bibinfo  {journal} {Nature}\ }\textbf {\bibinfo {volume} {590}},\ \bibinfo
  {pages} {238–242} (\bibinfo {year} {2021})}\BibitemShut {NoStop}%
\bibitem [{\citenamefont {Lee}\ \emph {et~al.}(2020)\citenamefont {Lee},
  \citenamefont {Ahn}, \citenamefont {Choi}, \citenamefont {Ko},\ and\
  \citenamefont {Semertzidis}}]{CAPPII}%
  \BibitemOpen
  \bibfield  {author} {\bibinfo {author} {\bibfnamefont {S.}~\bibnamefont
  {Lee}}, \bibinfo {author} {\bibfnamefont {S.}~\bibnamefont {Ahn}}, \bibinfo
  {author} {\bibfnamefont {J.}~\bibnamefont {Choi}}, \bibinfo {author}
  {\bibfnamefont {B.~R.}\ \bibnamefont {Ko}},\ and\ \bibinfo {author}
  {\bibfnamefont {Y.~K.}\ \bibnamefont {Semertzidis}},\ }\href
  {https://doi.org/10.1103/PhysRevLett.124.101802} {\bibfield  {journal}
  {\bibinfo  {journal} {Phys. Rev. Lett.}\ }\textbf {\bibinfo {volume} {124}},\
  \bibinfo {pages} {101802} (\bibinfo {year} {2020})}\BibitemShut {NoStop}%
\bibitem [{\citenamefont {Jeong}\ \emph {et~al.}(2020)\citenamefont {Jeong},
  \citenamefont {Youn}, \citenamefont {Bae}, \citenamefont {Kim}, \citenamefont
  {Seong}, \citenamefont {Kim},\ and\ \citenamefont {Semertzidis}}]{CAPPIII}%
  \BibitemOpen
  \bibfield  {author} {\bibinfo {author} {\bibfnamefont {J.}~\bibnamefont
  {Jeong}}, \bibinfo {author} {\bibfnamefont {S.~W.}\ \bibnamefont {Youn}},
  \bibinfo {author} {\bibfnamefont {S.}~\bibnamefont {Bae}}, \bibinfo {author}
  {\bibfnamefont {J.}~\bibnamefont {Kim}}, \bibinfo {author} {\bibfnamefont
  {T.}~\bibnamefont {Seong}}, \bibinfo {author} {\bibfnamefont {J.~E.}\
  \bibnamefont {Kim}},\ and\ \bibinfo {author} {\bibfnamefont {Y.~K.}\
  \bibnamefont {Semertzidis}},\ }\href
  {https://doi.org/10.1103/PhysRevLett.125.221302} {\bibfield  {journal}
  {\bibinfo  {journal} {Phys. Rev. Lett.}\ }\textbf {\bibinfo {volume} {125}},\
  \bibinfo {pages} {221302} (\bibinfo {year} {2020})}\BibitemShut {NoStop}%
\bibitem [{\citenamefont {Kwon}\ \emph {et~al.}(2021)\citenamefont {Kwon},
  \citenamefont {Lee}, \citenamefont {Chung}, \citenamefont {Ahn},
  \citenamefont {Byun}, \citenamefont {Caspers}, \citenamefont {Choi},
  \citenamefont {Choi}, \citenamefont {Chong}, \citenamefont {Jeong},
  \citenamefont {Jeong}, \citenamefont {Kim}, \citenamefont {Kim},
  \citenamefont {Kutlu}, \citenamefont {Lee}, \citenamefont {Lee},
  \citenamefont {Lee}, \citenamefont {Matlashov}, \citenamefont {Oh},
  \citenamefont {Park}, \citenamefont {Uchaikin}, \citenamefont {Youn},\ and\
  \citenamefont {Semertzidis}}]{CAPPI}%
  \BibitemOpen
  \bibfield  {author} {\bibinfo {author} {\bibfnamefont {O.}~\bibnamefont
  {Kwon}}, \bibinfo {author} {\bibfnamefont {D.}~\bibnamefont {Lee}}, \bibinfo
  {author} {\bibfnamefont {W.}~\bibnamefont {Chung}}, \bibinfo {author}
  {\bibfnamefont {D.}~\bibnamefont {Ahn}}, \bibinfo {author} {\bibfnamefont
  {H.~S.}\ \bibnamefont {Byun}}, \bibinfo {author} {\bibfnamefont
  {F.}~\bibnamefont {Caspers}}, \bibinfo {author} {\bibfnamefont
  {H.}~\bibnamefont {Choi}}, \bibinfo {author} {\bibfnamefont {J.}~\bibnamefont
  {Choi}}, \bibinfo {author} {\bibfnamefont {Y.}~\bibnamefont {Chong}},
  \bibinfo {author} {\bibfnamefont {H.}~\bibnamefont {Jeong}}, \bibinfo
  {author} {\bibfnamefont {J.}~\bibnamefont {Jeong}}, \bibinfo {author}
  {\bibfnamefont {J.~E.}\ \bibnamefont {Kim}}, \bibinfo {author} {\bibfnamefont
  {J.}~\bibnamefont {Kim}}, \bibinfo {author} {\bibfnamefont {C.}~\bibnamefont
  {Kutlu}}, \bibinfo {author} {\bibfnamefont {J.}~\bibnamefont {Lee}}, \bibinfo
  {author} {\bibfnamefont {M.~J.}\ \bibnamefont {Lee}}, \bibinfo {author}
  {\bibfnamefont {S.}~\bibnamefont {Lee}}, \bibinfo {author} {\bibfnamefont
  {A.}~\bibnamefont {Matlashov}}, \bibinfo {author} {\bibfnamefont
  {S.}~\bibnamefont {Oh}}, \bibinfo {author} {\bibfnamefont {S.}~\bibnamefont
  {Park}}, \bibinfo {author} {\bibfnamefont {S.}~\bibnamefont {Uchaikin}},
  \bibinfo {author} {\bibfnamefont {S.~W.}\ \bibnamefont {Youn}},\ and\
  \bibinfo {author} {\bibfnamefont {Y.~K.}\ \bibnamefont {Semertzidis}},\
  }\href {https://doi.org/10.1103/PhysRevLett.126.191802} {\bibfield  {journal}
  {\bibinfo  {journal} {Phys. Rev. Lett.}\ }\textbf {\bibinfo {volume} {126}},\
  \bibinfo {pages} {191802} (\bibinfo {year} {2021})}\BibitemShut {NoStop}%
\bibitem [{\citenamefont {Alesini}\ \emph {et~al.}(2021)\citenamefont
  {Alesini}, \citenamefont {Braggio}, \citenamefont {Carugno}, \citenamefont
  {Crescini}, \citenamefont {D'Agostino}, \citenamefont {Di~Gioacchino},
  \citenamefont {Di~Vora}, \citenamefont {Falferi}, \citenamefont
  {Gambardella}, \citenamefont {Gatti}, \citenamefont {Iannone}, \citenamefont
  {Ligi}, \citenamefont {Lombardi}, \citenamefont {Maccarrone}, \citenamefont
  {Ortolan}, \citenamefont {Pengo}, \citenamefont {Rettaroli}, \citenamefont
  {Ruoso}, \citenamefont {Taffarello},\ and\ \citenamefont {Tocci}}]{QUAX}%
  \BibitemOpen
  \bibfield  {author} {\bibinfo {author} {\bibfnamefont {D.}~\bibnamefont
  {Alesini}}, \bibinfo {author} {\bibfnamefont {C.}~\bibnamefont {Braggio}},
  \bibinfo {author} {\bibfnamefont {G.}~\bibnamefont {Carugno}}, \bibinfo
  {author} {\bibfnamefont {N.}~\bibnamefont {Crescini}}, \bibinfo {author}
  {\bibfnamefont {D.}~\bibnamefont {D'Agostino}}, \bibinfo {author}
  {\bibfnamefont {D.}~\bibnamefont {Di~Gioacchino}}, \bibinfo {author}
  {\bibfnamefont {R.}~\bibnamefont {Di~Vora}}, \bibinfo {author} {\bibfnamefont
  {P.}~\bibnamefont {Falferi}}, \bibinfo {author} {\bibfnamefont
  {U.}~\bibnamefont {Gambardella}}, \bibinfo {author} {\bibfnamefont
  {C.}~\bibnamefont {Gatti}}, \bibinfo {author} {\bibfnamefont
  {G.}~\bibnamefont {Iannone}}, \bibinfo {author} {\bibfnamefont
  {C.}~\bibnamefont {Ligi}}, \bibinfo {author} {\bibfnamefont {A.}~\bibnamefont
  {Lombardi}}, \bibinfo {author} {\bibfnamefont {G.}~\bibnamefont
  {Maccarrone}}, \bibinfo {author} {\bibfnamefont {A.}~\bibnamefont {Ortolan}},
  \bibinfo {author} {\bibfnamefont {R.}~\bibnamefont {Pengo}}, \bibinfo
  {author} {\bibfnamefont {A.}~\bibnamefont {Rettaroli}}, \bibinfo {author}
  {\bibfnamefont {G.}~\bibnamefont {Ruoso}}, \bibinfo {author} {\bibfnamefont
  {L.}~\bibnamefont {Taffarello}},\ and\ \bibinfo {author} {\bibfnamefont
  {S.}~\bibnamefont {Tocci}},\ }\href
  {https://doi.org/10.1103/PhysRevD.103.102004} {\bibfield  {journal} {\bibinfo
   {journal} {Phys. Rev. D}\ }\textbf {\bibinfo {volume} {103}},\ \bibinfo
  {pages} {102004} (\bibinfo {year} {2021})}\BibitemShut {NoStop}%
\bibitem [{\citenamefont {Choi}\ \emph {et~al.}(2021)\citenamefont {Choi},
  \citenamefont {Ahn}, \citenamefont {Ko}, \citenamefont {Lee},\ and\
  \citenamefont {Semertzidis}}]{CAPPCavity}%
  \BibitemOpen
  \bibfield  {author} {\bibinfo {author} {\bibfnamefont {J.}~\bibnamefont
  {Choi}}, \bibinfo {author} {\bibfnamefont {S.}~\bibnamefont {Ahn}}, \bibinfo
  {author} {\bibfnamefont {B.}~\bibnamefont {Ko}}, \bibinfo {author}
  {\bibfnamefont {S.}~\bibnamefont {Lee}},\ and\ \bibinfo {author}
  {\bibfnamefont {Y.}~\bibnamefont {Semertzidis}},\ }\href
  {https://doi.org/https://doi.org/10.1016/j.nima.2021.165667} {\bibfield
  {journal} {\bibinfo  {journal} {Nuclear Instruments and Methods in Physics
  Research Section A: Accelerators, Spectrometers, Detectors and Associated
  Equipment}\ }\textbf {\bibinfo {volume} {1013}},\ \bibinfo {pages} {165667}
  (\bibinfo {year} {2021})}\BibitemShut {NoStop}%
\bibitem [{\citenamefont {Chang}\ \emph
  {et~al.}(2022{\natexlab{a}})\citenamefont {Chang}, \citenamefont {Chang},
  \citenamefont {Chang}, \citenamefont {Chang}, \citenamefont {Chang},
  \citenamefont {Chen}, \citenamefont {Chen}, \citenamefont {Chen},
  \citenamefont {Chen}, \citenamefont {Chiang}, \citenamefont {Chien},
  \citenamefont {Doan}, \citenamefont {Hung}, \citenamefont {Kuo},
  \citenamefont {Lai}, \citenamefont {Liu}, \citenamefont {OuYang},
  \citenamefont {Wu},\ and\ \citenamefont {Yu}}]{TASEHInstrumentation}%
  \BibitemOpen
  \bibfield  {author} {\bibinfo {author} {\bibfnamefont {H.}~\bibnamefont
  {Chang}}, \bibinfo {author} {\bibfnamefont {J.-Y.}\ \bibnamefont {Chang}},
  \bibinfo {author} {\bibfnamefont {Y.-C.}\ \bibnamefont {Chang}}, \bibinfo
  {author} {\bibfnamefont {Y.-H.}\ \bibnamefont {Chang}}, \bibinfo {author}
  {\bibfnamefont {Y.-H.}\ \bibnamefont {Chang}}, \bibinfo {author}
  {\bibfnamefont {C.-H.}\ \bibnamefont {Chen}}, \bibinfo {author}
  {\bibfnamefont {C.-F.}\ \bibnamefont {Chen}}, \bibinfo {author}
  {\bibfnamefont {K.-Y.}\ \bibnamefont {Chen}}, \bibinfo {author}
  {\bibfnamefont {Y.-F.}\ \bibnamefont {Chen}}, \bibinfo {author}
  {\bibfnamefont {W.-Y.}\ \bibnamefont {Chiang}}, \bibinfo {author}
  {\bibfnamefont {W.-C.}\ \bibnamefont {Chien}}, \bibinfo {author}
  {\bibfnamefont {H.~T.}\ \bibnamefont {Doan}}, \bibinfo {author}
  {\bibfnamefont {W.-C.}\ \bibnamefont {Hung}}, \bibinfo {author}
  {\bibfnamefont {W.}~\bibnamefont {Kuo}}, \bibinfo {author} {\bibfnamefont
  {S.-B.}\ \bibnamefont {Lai}}, \bibinfo {author} {\bibfnamefont {H.-W.}\
  \bibnamefont {Liu}}, \bibinfo {author} {\bibfnamefont {M.-W.}\ \bibnamefont
  {OuYang}}, \bibinfo {author} {\bibfnamefont {P.-I.}\ \bibnamefont {Wu}},\
  and\ \bibinfo {author} {\bibfnamefont {S.-S.}\ \bibnamefont {Yu}} (\bibinfo
  {collaboration} {TASEH Collaboration}),\ }\href@noop {} {\  (\bibinfo {year}
  {2022}{\natexlab{a}})},\ \Eprint {https://arxiv.org/abs/2205.01477}
  {arXiv:2205.01477 [physics.ins-det]} \BibitemShut {NoStop}%
\bibitem [{\citenamefont {Alesini}\ \emph {et~al.}(2019)\citenamefont
  {Alesini}, \citenamefont {Braggio}, \citenamefont {Carugno}, \citenamefont
  {Crescini}, \citenamefont {D'Agostino}, \citenamefont {Di~Gioacchino},
  \citenamefont {Di~Vora}, \citenamefont {Falferi}, \citenamefont {Gallo},
  \citenamefont {Gambardella}, \citenamefont {Gatti}, \citenamefont {Iannone},
  \citenamefont {Lamanna}, \citenamefont {Ligi}, \citenamefont {Lombardi},
  \citenamefont {Mezzena}, \citenamefont {Ortolan}, \citenamefont {Pengo},
  \citenamefont {Pompeo}, \citenamefont {Rettaroli}, \citenamefont {Ruoso},
  \citenamefont {Silva}, \citenamefont {Speake}, \citenamefont {Taffarello},\
  and\ \citenamefont {Tocci}}]{AxionFormula}%
  \BibitemOpen
  \bibfield  {author} {\bibinfo {author} {\bibfnamefont {D.}~\bibnamefont
  {Alesini}}, \bibinfo {author} {\bibfnamefont {C.}~\bibnamefont {Braggio}},
  \bibinfo {author} {\bibfnamefont {G.}~\bibnamefont {Carugno}}, \bibinfo
  {author} {\bibfnamefont {N.}~\bibnamefont {Crescini}}, \bibinfo {author}
  {\bibfnamefont {D.}~\bibnamefont {D'Agostino}}, \bibinfo {author}
  {\bibfnamefont {D.}~\bibnamefont {Di~Gioacchino}}, \bibinfo {author}
  {\bibfnamefont {R.}~\bibnamefont {Di~Vora}}, \bibinfo {author} {\bibfnamefont
  {P.}~\bibnamefont {Falferi}}, \bibinfo {author} {\bibfnamefont
  {S.}~\bibnamefont {Gallo}}, \bibinfo {author} {\bibfnamefont
  {U.}~\bibnamefont {Gambardella}}, \bibinfo {author} {\bibfnamefont
  {C.}~\bibnamefont {Gatti}}, \bibinfo {author} {\bibfnamefont
  {G.}~\bibnamefont {Iannone}}, \bibinfo {author} {\bibfnamefont
  {G.}~\bibnamefont {Lamanna}}, \bibinfo {author} {\bibfnamefont
  {C.}~\bibnamefont {Ligi}}, \bibinfo {author} {\bibfnamefont {A.}~\bibnamefont
  {Lombardi}}, \bibinfo {author} {\bibfnamefont {R.}~\bibnamefont {Mezzena}},
  \bibinfo {author} {\bibfnamefont {A.}~\bibnamefont {Ortolan}}, \bibinfo
  {author} {\bibfnamefont {R.}~\bibnamefont {Pengo}}, \bibinfo {author}
  {\bibfnamefont {N.}~\bibnamefont {Pompeo}}, \bibinfo {author} {\bibfnamefont
  {A.}~\bibnamefont {Rettaroli}}, \bibinfo {author} {\bibfnamefont
  {G.}~\bibnamefont {Ruoso}}, \bibinfo {author} {\bibfnamefont
  {E.}~\bibnamefont {Silva}}, \bibinfo {author} {\bibfnamefont {C.~C.}\
  \bibnamefont {Speake}}, \bibinfo {author} {\bibfnamefont {L.}~\bibnamefont
  {Taffarello}},\ and\ \bibinfo {author} {\bibfnamefont {S.}~\bibnamefont
  {Tocci}},\ }\href {https://doi.org/10.1103/PhysRevD.99.101101} {\bibfield
  {journal} {\bibinfo  {journal} {Phys. Rev. D}\ }\textbf {\bibinfo {volume}
  {99}},\ \bibinfo {pages} {101101(R)} (\bibinfo {year} {2019})}\BibitemShut
  {NoStop}%
\bibitem [{\citenamefont {Read}(2014)}]{Read:2014qva}%
  \BibitemOpen
  \bibfield  {author} {\bibinfo {author} {\bibfnamefont {J.~I.}\ \bibnamefont
  {Read}},\ }\href {https://doi.org/10.1088/0954-3899/41/6/063101} {\bibfield
  {journal} {\bibinfo  {journal} {J. Phys. G: Nucl. Part. Phys.}\ }\textbf
  {\bibinfo {volume} {41}},\ \bibinfo {pages} {063101} (\bibinfo {year}
  {2014})}\BibitemShut {NoStop}%
\bibitem [{\citenamefont {Kim}(1979)}]{KSVZI}%
  \BibitemOpen
  \bibfield  {author} {\bibinfo {author} {\bibfnamefont {J.~E.}\ \bibnamefont
  {Kim}},\ }\href {https://doi.org/10.1103/PhysRevLett.43.103} {\bibfield
  {journal} {\bibinfo  {journal} {Phys. Rev. Lett.}\ }\textbf {\bibinfo
  {volume} {43}},\ \bibinfo {pages} {103} (\bibinfo {year} {1979})}\BibitemShut
  {NoStop}%
\bibitem [{\citenamefont {Shifman}\ \emph {et~al.}(1980)\citenamefont
  {Shifman}, \citenamefont {Vainshtein},\ and\ \citenamefont
  {Zakharov}}]{KSVZII}%
  \BibitemOpen
  \bibfield  {author} {\bibinfo {author} {\bibfnamefont {M.~A.}\ \bibnamefont
  {Shifman}}, \bibinfo {author} {\bibfnamefont {A.~I.}\ \bibnamefont
  {Vainshtein}},\ and\ \bibinfo {author} {\bibfnamefont {V.~I.}\ \bibnamefont
  {Zakharov}},\ }\href {https://doi.org/10.1016/0550-3213(80)90209-6}
  {\bibfield  {journal} {\bibinfo  {journal} {Nucl. Phys. B}\ }\textbf
  {\bibinfo {volume} {166}},\ \bibinfo {pages} {493} (\bibinfo {year}
  {1980})}\BibitemShut {NoStop}%
\bibitem [{\citenamefont {Dine}\ \emph {et~al.}(1981)\citenamefont {Dine},
  \citenamefont {Fischler},\ and\ \citenamefont {Srednicki}}]{DFSZI}%
  \BibitemOpen
  \bibfield  {author} {\bibinfo {author} {\bibfnamefont {M.}~\bibnamefont
  {Dine}}, \bibinfo {author} {\bibfnamefont {W.}~\bibnamefont {Fischler}},\
  and\ \bibinfo {author} {\bibfnamefont {M.}~\bibnamefont {Srednicki}},\ }\href
  {https://doi.org/10.1016/0370-2693(81)90590-6} {\bibfield  {journal}
  {\bibinfo  {journal} {Phys. Lett. B}\ }\textbf {\bibinfo {volume} {104}},\
  \bibinfo {pages} {199} (\bibinfo {year} {1981})}\BibitemShut {NoStop}%
\bibitem [{\citenamefont {Zhitnitsky}(1980)}]{DFSZII}%
  \BibitemOpen
  \bibfield  {author} {\bibinfo {author} {\bibfnamefont {A.~R.}\ \bibnamefont
  {Zhitnitsky}},\ }\href@noop {} {\bibfield  {journal} {\bibinfo  {journal}
  {Sov. J. Nucl. Phys.}\ }\textbf {\bibinfo {volume} {31}},\ \bibinfo {pages}
  {260} (\bibinfo {year} {1980})}\BibitemShut {NoStop}%
\bibitem [{\citenamefont {Dicke}(1946)}]{Dicke}%
  \BibitemOpen
  \bibfield  {author} {\bibinfo {author} {\bibfnamefont {R.~H.}\ \bibnamefont
  {Dicke}},\ }\href {https://doi.org/10.1063/1.1770483} {\bibfield  {journal}
  {\bibinfo  {journal} {Review of Scientific Instruments}\ }\textbf {\bibinfo
  {volume} {17}},\ \bibinfo {pages} {268} (\bibinfo {year} {1946})}\BibitemShut
  {NoStop}%
\bibitem [{\citenamefont {Chang}\ \emph
  {et~al.}(2022{\natexlab{b}})\citenamefont {Chang}, \citenamefont {Chang},
  \citenamefont {Chang}, \citenamefont {Chang}, \citenamefont {Chang},
  \citenamefont {Chen}, \citenamefont {Chen}, \citenamefont {Chen},
  \citenamefont {Chen}, \citenamefont {Chiang}, \citenamefont {Chien},
  \citenamefont {Doan}, \citenamefont {Hung}, \citenamefont {Kuo},
  \citenamefont {Lai}, \citenamefont {Liu}, \citenamefont {OuYang},
  \citenamefont {Wu},\ and\ \citenamefont {Yu}}]{TASEHAnalysis}%
  \BibitemOpen
  \bibfield  {author} {\bibinfo {author} {\bibfnamefont {H.}~\bibnamefont
  {Chang}}, \bibinfo {author} {\bibfnamefont {J.-Y.}\ \bibnamefont {Chang}},
  \bibinfo {author} {\bibfnamefont {Y.-C.}\ \bibnamefont {Chang}}, \bibinfo
  {author} {\bibfnamefont {Y.-H.}\ \bibnamefont {Chang}}, \bibinfo {author}
  {\bibfnamefont {Y.-H.}\ \bibnamefont {Chang}}, \bibinfo {author}
  {\bibfnamefont {C.-H.}\ \bibnamefont {Chen}}, \bibinfo {author}
  {\bibfnamefont {C.-F.}\ \bibnamefont {Chen}}, \bibinfo {author}
  {\bibfnamefont {K.-Y.}\ \bibnamefont {Chen}}, \bibinfo {author}
  {\bibfnamefont {Y.-F.}\ \bibnamefont {Chen}}, \bibinfo {author}
  {\bibfnamefont {W.-Y.}\ \bibnamefont {Chiang}}, \bibinfo {author}
  {\bibfnamefont {W.-C.}\ \bibnamefont {Chien}}, \bibinfo {author}
  {\bibfnamefont {H.~T.}\ \bibnamefont {Doan}}, \bibinfo {author}
  {\bibfnamefont {W.-C.}\ \bibnamefont {Hung}}, \bibinfo {author}
  {\bibfnamefont {W.}~\bibnamefont {Kuo}}, \bibinfo {author} {\bibfnamefont
  {S.-B.}\ \bibnamefont {Lai}}, \bibinfo {author} {\bibfnamefont {H.-W.}\
  \bibnamefont {Liu}}, \bibinfo {author} {\bibfnamefont {M.-W.}\ \bibnamefont
  {OuYang}}, \bibinfo {author} {\bibfnamefont {P.-I.}\ \bibnamefont {Wu}},\
  and\ \bibinfo {author} {\bibfnamefont {S.-S.}\ \bibnamefont {Yu}} (\bibinfo
  {collaboration} {TASEH Collaboration}),\ }\href@noop {} {\  (\bibinfo {year}
  {2022}{\natexlab{b}})},\ \Eprint {https://arxiv.org/abs/2204.14265}
  {arXiv:2204.14265 [hep-ex]} \BibitemShut {NoStop}%
\bibitem [{\citenamefont {Brubaker}\ \emph
  {et~al.}(2017{\natexlab{b}})\citenamefont {Brubaker}, \citenamefont {Zhong},
  \citenamefont {Lamoreaux}, \citenamefont {Lehnert},\ and\ \citenamefont {van
  Bibber}}]{HAYSTACII}%
  \BibitemOpen
  \bibfield  {author} {\bibinfo {author} {\bibfnamefont {B.~M.}\ \bibnamefont
  {Brubaker}}, \bibinfo {author} {\bibfnamefont {L.}~\bibnamefont {Zhong}},
  \bibinfo {author} {\bibfnamefont {S.~K.}\ \bibnamefont {Lamoreaux}}, \bibinfo
  {author} {\bibfnamefont {K.~W.}\ \bibnamefont {Lehnert}},\ and\ \bibinfo
  {author} {\bibfnamefont {K.~A.}\ \bibnamefont {van Bibber}},\ }\href
  {https://doi.org/10.1103/physrevd.96.123008} {\bibfield  {journal} {\bibinfo
  {journal} {Phys. Rev. D}\ }\textbf {\bibinfo {volume} {96}},\ \bibinfo
  {pages} {123008} (\bibinfo {year} {2017}{\natexlab{b}})}\BibitemShut
  {NoStop}%
\bibitem [{\citenamefont {Savitzky}\ and\ \citenamefont
  {Golay}(1964)}]{SGFilter}%
  \BibitemOpen
  \bibfield  {author} {\bibinfo {author} {\bibfnamefont {A.}~\bibnamefont
  {Savitzky}}\ and\ \bibinfo {author} {\bibfnamefont {M.~J.~E.}\ \bibnamefont
  {Golay}},\ }\href {https://doi.org/10.1021/ac60214a047} {\bibfield  {journal}
  {\bibinfo  {journal} {Anal. Chem.}\ }\textbf {\bibinfo {volume} {36}},\
  \bibinfo {pages} {1627} (\bibinfo {year} {1964})}\BibitemShut {NoStop}%
\bibitem [{\citenamefont {Bartram}\ \emph
  {et~al.}(2021{\natexlab{b}})\citenamefont {Bartram}, \citenamefont {Braine},
  \citenamefont {Cervantes}, \citenamefont {Crisosto}, \citenamefont {Du},
  \citenamefont {Leum}, \citenamefont {Rosenberg}, \citenamefont {Rybka},
  \citenamefont {Yang}, \citenamefont {Bowring}, \citenamefont {Chou},
  \citenamefont {Khatiwada}, \citenamefont {Sonnenschein}, \citenamefont
  {Wester}, \citenamefont {Carosi}, \citenamefont {Woollett}, \citenamefont
  {Duffy}, \citenamefont {Goryachev}, \citenamefont {McAllister}, \citenamefont
  {Tobar}, \citenamefont {Boutan}, \citenamefont {Jones}, \citenamefont
  {LaRoque}, \citenamefont {Oblath}, \citenamefont {Taubman}, \citenamefont
  {Clarke}, \citenamefont {Dove}, \citenamefont {Eddins}, \citenamefont
  {O'Kelley}, \citenamefont {Nawaz}, \citenamefont {Siddiqi}, \citenamefont
  {Stevenson}, \citenamefont {Agrawal}, \citenamefont {Dixit}, \citenamefont
  {Gleason}, \citenamefont {Jois}, \citenamefont {Sikivie}, \citenamefont
  {Solomon}, \citenamefont {Sullivan}, \citenamefont {Tanner}, \citenamefont
  {Lentz}, \citenamefont {Daw}, \citenamefont {Perry}, \citenamefont {Buckley},
  \citenamefont {Harrington}, \citenamefont {Henriksen},\ and\ \citenamefont
  {Murch}}]{ADMXVIII}%
  \BibitemOpen
  \bibfield  {author} {\bibinfo {author} {\bibfnamefont {C.}~\bibnamefont
  {Bartram}}, \bibinfo {author} {\bibfnamefont {T.}~\bibnamefont {Braine}},
  \bibinfo {author} {\bibfnamefont {R.}~\bibnamefont {Cervantes}}, \bibinfo
  {author} {\bibfnamefont {N.}~\bibnamefont {Crisosto}}, \bibinfo {author}
  {\bibfnamefont {N.}~\bibnamefont {Du}}, \bibinfo {author} {\bibfnamefont
  {G.}~\bibnamefont {Leum}}, \bibinfo {author} {\bibfnamefont {L.~J.}\
  \bibnamefont {Rosenberg}}, \bibinfo {author} {\bibfnamefont {G.}~\bibnamefont
  {Rybka}}, \bibinfo {author} {\bibfnamefont {J.}~\bibnamefont {Yang}},
  \bibinfo {author} {\bibfnamefont {D.}~\bibnamefont {Bowring}}, \bibinfo
  {author} {\bibfnamefont {A.~S.}\ \bibnamefont {Chou}}, \bibinfo {author}
  {\bibfnamefont {R.}~\bibnamefont {Khatiwada}}, \bibinfo {author}
  {\bibfnamefont {A.}~\bibnamefont {Sonnenschein}}, \bibinfo {author}
  {\bibfnamefont {W.}~\bibnamefont {Wester}}, \bibinfo {author} {\bibfnamefont
  {G.}~\bibnamefont {Carosi}}, \bibinfo {author} {\bibfnamefont
  {N.}~\bibnamefont {Woollett}}, \bibinfo {author} {\bibfnamefont {L.~D.}\
  \bibnamefont {Duffy}}, \bibinfo {author} {\bibfnamefont {M.}~\bibnamefont
  {Goryachev}}, \bibinfo {author} {\bibfnamefont {B.}~\bibnamefont
  {McAllister}}, \bibinfo {author} {\bibfnamefont {M.~E.}\ \bibnamefont
  {Tobar}}, \bibinfo {author} {\bibfnamefont {C.}~\bibnamefont {Boutan}},
  \bibinfo {author} {\bibfnamefont {M.}~\bibnamefont {Jones}}, \bibinfo
  {author} {\bibfnamefont {B.~H.}\ \bibnamefont {LaRoque}}, \bibinfo {author}
  {\bibfnamefont {N.~S.}\ \bibnamefont {Oblath}}, \bibinfo {author}
  {\bibfnamefont {M.~S.}\ \bibnamefont {Taubman}}, \bibinfo {author}
  {\bibfnamefont {J.}~\bibnamefont {Clarke}}, \bibinfo {author} {\bibfnamefont
  {A.}~\bibnamefont {Dove}}, \bibinfo {author} {\bibfnamefont {A.}~\bibnamefont
  {Eddins}}, \bibinfo {author} {\bibfnamefont {S.~R.}\ \bibnamefont
  {O'Kelley}}, \bibinfo {author} {\bibfnamefont {S.}~\bibnamefont {Nawaz}},
  \bibinfo {author} {\bibfnamefont {I.}~\bibnamefont {Siddiqi}}, \bibinfo
  {author} {\bibfnamefont {N.}~\bibnamefont {Stevenson}}, \bibinfo {author}
  {\bibfnamefont {A.}~\bibnamefont {Agrawal}}, \bibinfo {author} {\bibfnamefont
  {A.~V.}\ \bibnamefont {Dixit}}, \bibinfo {author} {\bibfnamefont {J.~R.}\
  \bibnamefont {Gleason}}, \bibinfo {author} {\bibfnamefont {S.}~\bibnamefont
  {Jois}}, \bibinfo {author} {\bibfnamefont {P.}~\bibnamefont {Sikivie}},
  \bibinfo {author} {\bibfnamefont {J.~A.}\ \bibnamefont {Solomon}}, \bibinfo
  {author} {\bibfnamefont {N.~S.}\ \bibnamefont {Sullivan}}, \bibinfo {author}
  {\bibfnamefont {D.~B.}\ \bibnamefont {Tanner}}, \bibinfo {author}
  {\bibfnamefont {E.}~\bibnamefont {Lentz}}, \bibinfo {author} {\bibfnamefont
  {E.~J.}\ \bibnamefont {Daw}}, \bibinfo {author} {\bibfnamefont {M.~G.}\
  \bibnamefont {Perry}}, \bibinfo {author} {\bibfnamefont {J.~H.}\ \bibnamefont
  {Buckley}}, \bibinfo {author} {\bibfnamefont {P.~M.}\ \bibnamefont
  {Harrington}}, \bibinfo {author} {\bibfnamefont {E.~A.}\ \bibnamefont
  {Henriksen}},\ and\ \bibinfo {author} {\bibfnamefont {K.~W.}\ \bibnamefont
  {Murch}} (\bibinfo {collaboration} {ADMX Collaboration}),\ }\href
  {https://doi.org/10.1103/PhysRevD.103.032002} {\bibfield  {journal} {\bibinfo
   {journal} {Phys. Rev. D}\ }\textbf {\bibinfo {volume} {103}},\ \bibinfo
  {pages} {032002} (\bibinfo {year} {2021}{\natexlab{b}})}\BibitemShut
  {NoStop}%
\end{thebibliography}
\end{document}